# Bridging quantum mechanics and nonlinear optics in Raman scattering

Naoki Fukutake[1,2*] and Hideaki Kano[3]

[1]*Nikon Corporation, 1-5-20 Nishioi, Shinagawa, Tokyo 140-8601, Japan.*
[2]*Institute of Pure and Applied Science, University of Tsukuba, Tennodai 1-1-1, Tsukuba, Ibaraki 305-8573 Japan.*
[3]*Department of Biosciences and Informatics, Faculty of Science and Technology, Keio University,3-14-1 Hiyoshi, Kohoku-ku, Yokohama, Kanagawa 223-8522, Japan.*

*Naoki.Fukutake@nikon.com

## ABSTRACT

We present a theoretical framework for spontaneous Raman scattering that fundamentally bridges quantum-mechanical and nonlinear-optical approaches. By conceptualizing spontaneous Raman scattering as a stimulated Raman gain or loss event seeded by the quantum vacuum field, we rigorously derive the spontaneous Raman cross-section directly from the third-order nonlinear susceptibility. Crucially, this framework predicts the existence of a hitherto unrecognized phenomenon: "spontaneous Raman loss" (sRL), which acts as the vacuum-seeded counterpart to stimulated Raman loss, complementing traditional spontaneous Raman scattering (spontaneous Raman gain, sRG). Furthermore, we establish a rigorous connection to the traditional Kramers-Heisenberg-Dirac (KHD) theory, revealing that the spontaneous process is governed by interference before a detector between the signal field emitted from molecules and the vacuum field itself that stimulates the molecules. This insight uncovers a direct correspondence between the sRG susceptibility and the rotating/counter-rotating interference terms in the KHD formula. Ultimately, we extend the foundational KHD theory by incorporating previously unrecognized essential terms, achieving perfect analytical agreement between the quantum mechanical and nonlinear optical descriptions of Raman scattering.

## I. INTRODUCTION

Raman spectroscopy has become an indispensable analytical tool in a wide range of fields, from material science to life sciences, serving as a non-destructive technique that provides detailed information on the vibrational structure of molecules. The theoretical foundation of the Raman effect was established even prior to its experimental discovery [1] in 1928. In 1925, the formula proposed by Kramers and Heisenberg [2], and later extended by Dirac [3]—known as the Kramers-Heisenberg-Dirac (KHD) dispersion formula—provided a universal theoretical formalism to describe Raman scattering in strict quantum mechanical terms.

The core of this theory lies in describing the light-molecule interaction using perturbation theory. Within this framework, light scattering is modeled as a second-order process involving the absorption of a single photon followed by the emission of a new photon. Specifically, a molecule transitions from an initial state to a virtual intermediate state by absorbing a photon and immediately proceeds to a final state by emitting another photon. This intermediate state is an extremely short-lived, non-stationary state governed by the time-energy uncertainty principle. The KHD formula quantitatively provides the transition probability for this entire process, from which the Raman scattering cross-section is directly derived. Notably, it inherently accounts for the resonance Raman effect [4], where the scattering cross section increases dramatically as the incident photon energy approaches a molecular electronic excitation energy, causing the denominator in the formula to approach zero.

Raman scattering is known to exhibit two distinct aspects: a spontaneous process and a stimulated one. While stimulated Raman scattering (SRS) is understood as a macroscopic nonlinear amplification process involving both pump and Stokes fields, spontaneous Raman scattering has traditionally been described as an inelastic scattering event from a single pump beam. However, these two phenomena can be described within a unified framework by conceptualizing spontaneous Raman scattering as an SRS process seeded by the quantum vacuum field [5−7]. In this picture, the quantum fluctuations of the vacuum field at the Stokes frequency act as a seed to generate real Stokes photons from the vacuum. It has also been implied that the spontaneous process is expressed by interference at a detector between the signal field emitted from molecules and the seed vacuum field [5,6]. Therefore, spontaneous Raman scattering is essentially a stimulated emission process initiated by the vacuum and is observed through the interference between the emitted field and the vacuum field itself.

From this perspective, the cross-sections of spontaneous Raman scattering and SRS—which are conventionally calculated separately—can be directly connected through a common macroscopic physical quantity: the third-order nonlinear susceptibility $\chi^{(3)}$. In this paper, we establish this unified theoretical framework and demonstrate how to rigorously derive the spontaneous Raman scattering cross section directly from the nonlinear susceptibility. A profound consequence of our approach is the theoretical prediction of a novel phenomenon, which we term "spontaneous Raman loss" (sRL), where pump photons are annihilated through a vacuum-seeded spontaneous process. Finally, we uncover the exact physical connection between our macroscopic nonlinear optical approach and the microscopic KHD theory. We demonstrate that by extending the traditional KHD formula—incorporating essential terms—the quantum mechanical description achieves perfect analytical agreement with the cross section derived from the nonlinear susceptibility.

## II. RESULTS

We calculate the intensity of spontaneous Raman scattering in two different ways: (A) using KHD theory and (B) using nonlinear optical susceptibility.

### A. Traditional KHD theory

According to the KHD dispersion formula [3,8−10], the differential cross section of spontaneous Raman scattering $d\sigma/d\Omega$ is given by

$$\frac{d\sigma_{\mathrm{KHD}}}{d\Omega} = \frac{N}{16\pi^2 {\varepsilon_{\mathrm{o}}}^2 c^4} \frac{\omega_{\mathrm{i}}(\omega_{\mathrm{i}} - \omega_{\mathrm{vib}})^3}{\hbar^2} \left| \sum_{\mathrm{e}} \frac{\langle \mathrm{v}|\mu|\mathrm{e}\rangle\langle \mathrm{e}|\mu|\mathrm{g}\rangle}{\omega_{\mathrm{eg}} - \omega_{\mathrm{i}} - i\Gamma} + \sum_{\mathrm{e}} \frac{\langle \mathrm{v}|\mu|\mathrm{e}\rangle\langle \mathrm{e}|\mu|\mathrm{g}\rangle}{\omega_{\mathrm{ev}} + \omega_{\mathrm{i}} - i\Gamma} \right|^2, \quad (1)$$

where $\sigma_{\mathrm{KHD}}$ represents the total cross section and $d\Omega$ represents an infinitesimal solid angle; $N$ stands for the molecular density; $\varepsilon_{\mathrm{o}}$ is the vacuum permittivity; $\omega_{\mathrm{vib}} = \omega_{\mathrm{i}} - \omega_{\mathrm{s}}$ is the vibrational frequency of the target molecule; $\omega_{\mathrm{eg}}$ denotes the energy-level difference between the ground state $|\mathrm{g}\rangle$ and electronically excited state $|\mathrm{e}\rangle$; $\langle \mathrm{e}|\mu|\mathrm{g}\rangle$ and $\langle \mathrm{v}|\mu|\mathrm{e}\rangle$ represent the transition dipole moments from the ground state $|\mathrm{g}\rangle$ to the electronically excited state $|\mathrm{e}\rangle$ and from the electronically excited state $|\mathrm{e}\rangle$ to the vibrationally excited state (in electronically ground state) $|\mathrm{v}\rangle$, respectively; $\omega_{\mathrm{ev}}$ stands for the energy-level difference between the states $|\mathrm{g}\rangle$ and $|\mathrm{v}\rangle$; and $\Gamma$ is a damping factor. Regarding the damping factors $\Gamma$ in the two terms in square of modulus in Eq. (1), we apply negative signs for both factors to be consistent with the later discussion (DISCUSSION A).

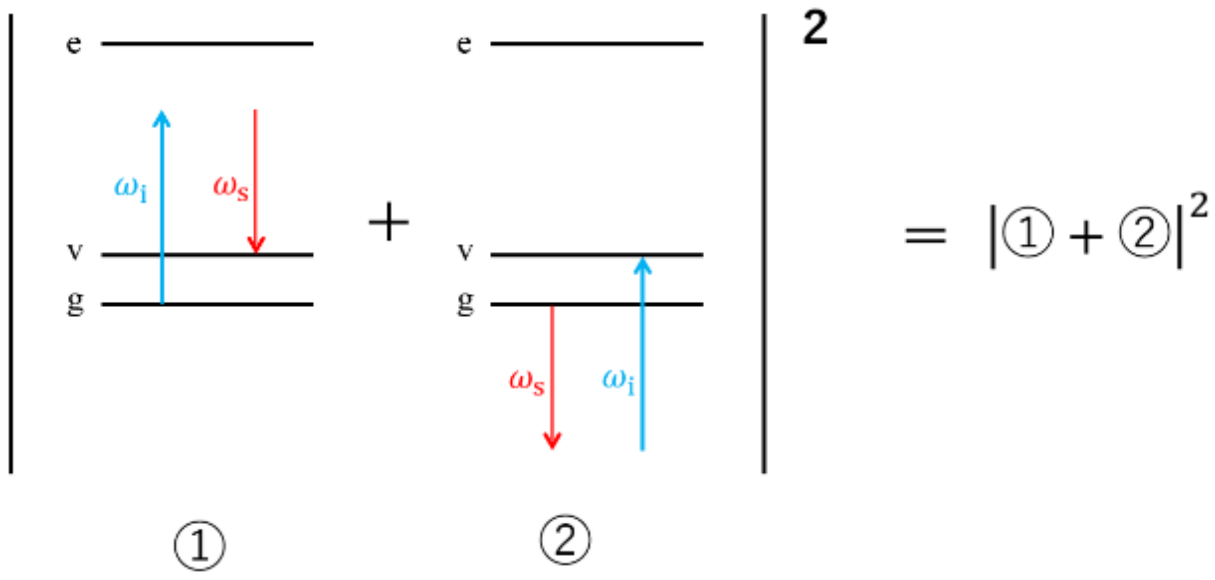

Fig. 1. Two types of processes involved in KHD theory: ① (rotating term) a process that passes through an intermediate state where one incident photon is absorbed, and ② (counter-rotating term) a process that passes through an intermediate state where one scattered photon is emitted.

Figure 1 describes the first term (rotating term) ① and the second term (counter-rotating term) ② in the square of the modulus in Eq. (1). The first term is much larger than the second term. For the later discussion (DISCUSSION A), we expand the square of the modulus, where Eq. (1) is decomposed into four terms, i.e., a large term ($T_1$), two small terms (cross terms $T_2$ and $T_3$), and a negligible term ($T_4$), as presented below.

$$T_1 = \frac{N}{16\pi^2{\varepsilon_{\mathrm{o}}}^2 c^4}\frac{\omega_{\mathrm{i}}(\omega_{\mathrm{i}}-\omega_{\mathrm{vib}})^3}{\hbar^2}\left|\sum_{\mathrm{e}}\frac{\langle\mathrm{v}|\mu|\mathrm{e}\rangle\langle\mathrm{e}|\mu|\mathrm{g}\rangle}{\omega_{\mathrm{eg}}-\omega_{\mathrm{i}}-i\Gamma}\right|^2 \tag{2}$$

$$T_2 = \frac{N}{16\pi^2{\varepsilon_{\mathrm{o}}}^2 c^4}\frac{\omega_{\mathrm{i}}(\omega_{\mathrm{i}}-\omega_{\mathrm{vib}})^3}{\hbar^2}\left(\sum_{\mathrm{e}}\frac{\langle\mathrm{v}|\mu|\mathrm{e}\rangle\langle\mathrm{e}|\mu|\mathrm{g}\rangle}{\omega_{\mathrm{eg}}-\omega_{\mathrm{i}}-i\Gamma}\right)\left(\sum_{\mathrm{e}}\frac{\langle\mathrm{v}|\mu|\mathrm{e}\rangle\langle\mathrm{e}|\mu|\mathrm{g}\rangle}{\omega_{\mathrm{ev}}+\omega_{\mathrm{i}}-i\Gamma}\right)^* \tag{3}$$

$$T_3 = \frac{N}{16\pi^2{\varepsilon_{\mathrm{o}}}^2 c^4}\frac{\omega_{\mathrm{i}}(\omega_{\mathrm{i}}-\omega_{\mathrm{vib}})^3}{\hbar^2}\left(\sum_{\mathrm{e}}\frac{\langle\mathrm{v}|\mu|\mathrm{e}\rangle\langle\mathrm{e}|\mu|\mathrm{g}\rangle}{\omega_{\mathrm{eg}}-\omega_{\mathrm{i}}-i\Gamma}\right)^*\left(\sum_{\mathrm{e}}\frac{\langle\mathrm{v}|\mu|\mathrm{e}\rangle\langle\mathrm{e}|\mu|\mathrm{g}\rangle}{\omega_{\mathrm{ev}}+\omega_{\mathrm{i}}-i\Gamma}\right) \tag{4}$$

$$T_4 = \frac{N}{16\pi^2{\varepsilon_{\mathrm{o}}}^2 c^4}\frac{\omega_{\mathrm{i}}(\omega_{\mathrm{i}}-\omega_{\mathrm{vib}})^3}{\hbar^2}\left|\sum_{\mathrm{e}}\frac{\langle\mathrm{v}|\mu|\mathrm{e}\rangle\langle\mathrm{e}|\mu|\mathrm{g}\rangle}{\omega_{\mathrm{ev}}+\omega_{\mathrm{i}}-i\Gamma}\right|^2 \tag{5}$$

## B. Nonlinear optical susceptibility approach incorporating the vacuum field

To include the vacuum field in nonlinear optics, we quantize all fields. First, we consider the stimulated Raman gain (SRG) and stimulated Raman loss (SRL) [11], where the pump field $\hat{E}_{\mathrm{i}}$ and Stokes field $\hat{E}_{\mathrm{s}}$ described below are applied to the target molecules.

$$\hat{E}_{\mathrm{i}} = \sqrt{\frac{\hbar\omega_{\mathrm{i}}}{2\varepsilon_{\mathrm{o}}\mathcal{V}}}\hat{a}_{\mathrm{i}}, \quad \hat{E}_{\mathrm{i}}^\dagger = \sqrt{\frac{\hbar\omega_{\mathrm{i}}}{2\varepsilon_{\mathrm{o}}\mathcal{V}}}\hat{a}_{\mathrm{i}}^\dagger \tag{6}$$

$$\hat{E}_{\mathrm{s}} = \sqrt{\frac{\hbar\omega_{\mathrm{s}}}{2\varepsilon_{\mathrm{o}}\mathcal{V}}}\hat{a}_{\mathrm{s}}, \quad \hat{E}_{\mathrm{s}}^\dagger = \sqrt{\frac{\hbar\omega_{\mathrm{i}}}{2\varepsilon_{\mathrm{o}}\mathcal{V}}}\hat{a}_{\mathrm{s}}^\dagger \tag{7}$$

Here $\mathcal{V}\to\infty$ represents a spatial volume, which eventually cancels out and disappears from the equations; $\hat{a}_{\mathrm{i}}$ is the annihilation operator of the pump field; and $\hat{a}_{\mathrm{s}}$ is the annihilation operator of the Stokes field. Considering the phase shift $-i$ between the signal field and the pump field (Stokes field), stemmed from the Gouy phase shift of a secondary wave, SRG and SRL are expressed as follows [5,6]:

$$\text{SRG:}\quad \left|-i\hat{E}_\mathrm{s}+\chi_\mathrm{SRG}^{(3)}\hat{E}_\mathrm{i}\hat{E}_\mathrm{s}\hat{E}_\mathrm{i}^\dagger\right|^2 \approx \hat{E}_\mathrm{s}^\dagger\hat{E}_\mathrm{s} - 2\mathrm{Im}\left\{\chi_\mathrm{SRG}^{(3)}\right\}\hat{E}_\mathrm{i}^\dagger\hat{E}_\mathrm{i}\hat{E}_\mathrm{s}^\dagger\hat{E}_\mathrm{s}, \tag{8}$$

$$\text{SRL:}\quad \left|-iE_\mathrm{i}+\chi_\mathrm{SRL}^{(3)}\hat{E}_\mathrm{s}\hat{E}_\mathrm{s}^\dagger\hat{E}_\mathrm{i}\right|^2 \approx \hat{E}_\mathrm{i}^\dagger\hat{E}_\mathrm{i} - 2\mathrm{Im}\left\{\chi_\mathrm{SRL}^{(3)}\right\}\hat{E}_\mathrm{i}^\dagger\hat{E}_\mathrm{i}\hat{E}_\mathrm{s}^\dagger\hat{E}_\mathrm{s}. \tag{9}$$

Here $\chi_\mathrm{SRG}^{(3)}$ and $\chi_\mathrm{SRL}^{(3)}$ indicate the nonlinear susceptibilities for SRG and SRL, respectively. Note that $-2\mathrm{Im}\left\{\chi_\mathrm{SRG}^{(3)}\right\}$ is a positive value and $-2\mathrm{Im}\left\{\chi_\mathrm{SRL}^{(3)}\right\}$ is a negative value, but both the absolute values are identical; furthermore, SRG and SRL always occur together.

In the typical experiments on SRG or SRL, laser beams are utilized for the pump field $\hat{E}_\mathrm{i}$ and Stokes fields $\hat{E}_\mathrm{s}$. Therefore, coherent states are assumed for the pump and Stokes fields in the following derivation, but the final formula of the differential cross section $d\sigma/d\Omega$ holds true also for number states.

The coherent states for the pump field $|\alpha_\mathrm{i}\rangle$ and Stokes field $|\alpha_\mathrm{s}\rangle$ have the following relations with the operators:

$$\hat{a}_\mathrm{i}|\alpha_\mathrm{i}\rangle = \alpha_\mathrm{i}|\alpha_\mathrm{i}\rangle, \tag{10}$$

$$\hat{a}_\mathrm{s}|\alpha_\mathrm{s}\rangle = \alpha_\mathrm{s}|\alpha_\mathrm{s}\rangle. \tag{11}$$

Here $\alpha_\mathrm{i}$ and $\alpha_\mathrm{s}$ are the eigenvalues of the annihilation operators $\hat{a}_\mathrm{i}$ and $\hat{a}_\mathrm{s}$, respectively. By sandwiching Eqs. (8) and (9) with $\langle\alpha_\mathrm{s}|\langle\alpha_\mathrm{i}|\cdots|\alpha_\mathrm{i}\rangle|\alpha_\mathrm{s}\rangle$, the expectation values for SRG and SRL are obtained:

$$\text{SRG:}\quad |E_\mathrm{s}|^2 - 2\mathrm{Im}\left\{\chi_\mathrm{SRG}^{(3)}\right\}|E_\mathrm{i}|^2|E_\mathrm{s}|^2, \tag{12}$$

$$\text{SRL:}\quad |E_\mathrm{i}|^2 - 2\mathrm{Im}\left\{\chi_\mathrm{SRL}^{(3)}\right\}|E_\mathrm{i}|^2|E_\mathrm{s}|^2, \tag{13}$$

with

$$|E_\mathrm{i}|^2 = \frac{|\alpha_\mathrm{i}|^2\hbar\omega_\mathrm{i}}{2\varepsilon_\mathrm{o}\mathcal{V}}, \tag{14}$$

$$|E_\mathrm{s}|^2 = \frac{|\alpha_\mathrm{s}|^2\hbar\omega_\mathrm{s}}{2\varepsilon_\mathrm{o}\mathcal{V}}. \tag{15}$$

In most experiments of SRG or SRL, the second terms in Eq. (12) for SRG or Eq. (13) for SRL are extracted with a lock-in detection method, etc.

Next, we deal with spontaneous Raman scattering, where only the pump beam from a laser source is applied to the target molecules and no Stokes field source is provided. While SRG and SRL are the optical processes stimulated by the Stokes field, spontaneous Raman scattering can be considered an optical process induced by the vacuum field. For the Stokes field in spontaneous Raman scattering, the coherent state $|\alpha_\mathrm{s}\rangle$ is replaced with the vacuum state $|0_\mathrm{s}\rangle$. Here we define *spontaneous* Raman gain (sRG) and sRL (*spontaneous* Raman loss) as the SRG and SRL induced by the vacuum field, respectively. To calculate the expected values for the sRG and sRL, we sandwich Eqs. (8) and (9) with $\langle 0_\mathrm{s}|\langle\alpha_\mathrm{i}|\cdots|\alpha_\mathrm{i}\rangle|0_\mathrm{s}\rangle$, resulting in

$$\text{sRL:}\quad |E_\mathrm{i}|^2 - 2\mathrm{Im}\left\{\chi_\mathrm{SRL}^{(3)}\right\}|E_\mathrm{i}|^2|V_\mathrm{s}|^2, \tag{16}$$

$$\text{sRG:}\quad |V_\mathrm{s}|^2 - 2\mathrm{Im}\left\{\chi_\mathrm{SRG}^{(3)}\right\}|E_\mathrm{i}|^2|V_\mathrm{s}|^2, \tag{17}$$

with

$$|V_\mathrm{s}|^2 = \frac{\hbar\omega_\mathrm{s}}{2\varepsilon_\mathrm{o}\mathcal{V}}\,. \tag{18}$$

In the calculation, we reordered the operators such that the creation operator is on the left for the real field and annihilation operator is on the left for the vacuum field, and used the relation

$$\hat{a}_\mathrm{s}^\dagger|0_\mathrm{s}\rangle = |1_\mathrm{s}\rangle. \tag{19}$$

The first term of sRG in Eq. (17) corresponds to the vacuum itself and cannot be observed. Although the terms $\left|\chi_{\rm SRL}^{(3)}\right|^2|E_{\rm i}|^4|V_{\rm s}|^2$ in sRL and $\left|\chi_{\rm SRG}^{(3)}\right|^2|E_{\rm i}|^4|V_{\rm s}|^2$ in sRG are observable, their amounts are negligible. The second terms in sRL and sRG provide the observed intensities related to the spontaneous Raman scattering $\Delta I_{\rm sRL}$ and $\Delta I_{\rm sRG}$, respectively:

$$\Delta I_{\rm sRL} = -2{\rm Im}\left\{\chi_{\rm SRL}^{(3)}\right\}|E_{\rm i}|^2|V_{\rm s}|^2, \tag{20}$$

$$\Delta I_{\rm sRG} = -2{\rm Im}\left\{\chi_{\rm SRG}^{(3)}\right\}|E_{\rm i}|^2|V_{\rm s}|^2. \tag{21}$$

In nonresonant Raman scattering, $\Delta I_{\rm sRG}$ has a positive value and $\Delta I_{\rm sRL}$ indicates a negative value. In this case, $\Delta I_{\rm sRL}$ corresponds to the amount decreased from the pump beam, i.e., the reduced amount in sRL is transformed into $\Delta I_{\rm sRG}$. Note that both the spontaneous optical processes, i.e., sRG and sRL, always occur simultaneously.

$\Delta I_{\rm sRL}$ and $\Delta I_{\rm sRG}$ corresponds to the detection intensity proportional to the number of photons created in spontaneous Raman scattering $n_{\rm G}$ and the number of photons that annihilate from the pump beam $n_{\rm L}$, respectively, where $n_{\rm G} = n_{\rm L}(= n)$. Namely, $\Delta I_{\rm sRL}$ and $\Delta I_{\rm sRG}$ have properties that their transition rates for sRG and sRL in unit time are identical. It is also noticeable that the same number of molecular-vibration quanta with an energy of $\hbar\omega_{\rm vib}$ are generated simultaneously: $n\hbar\omega_{\rm i} = n\hbar\omega_{\rm s} + n\hbar\omega_{\rm vib}$.

Here, we derive the photon numbers generated in spontaneous Raman scattering. The energy of spontaneous Raman-scattered light within a Cartesian coordinate microvolume $\Delta x\Delta y\Delta z$ can be defined as $\Delta U_{\rm C}\ [{\rm J}] = U\ [{\rm J/m^3}] \times \Delta x\Delta y\Delta z\ [{\rm m^3}]$, where $U = \varepsilon_0\Delta I_{\rm sRG}$. The energy density at a distance $r$ from the target molecules can be written as

$$U' = U\frac{\Delta x\Delta y\Delta z}{r^2 dr}\ [{\rm J/m^3}]. \tag{22}$$

The energy within a polar coordinate microvolume $r^2 dr d\Omega$ at the position mentioned above is

$$\Delta U_{\rm p} = U\frac{\Delta x\Delta y\Delta z}{r^2 dr} \times r^2 dr d\Omega = U\Delta x\Delta y\Delta z d\Omega. \tag{23}$$

Assuming that $\Delta x\Delta y\Delta z = 1$, we obtain $\Delta U_{\rm p} = U d\Omega$, which has the dimension of energy. Defining the transition speed as $1/\Delta\tau$, the number of photons generated in spontaneous Raman scattering in a time $\Delta\tau$ is given by

$$\Delta n_{\rm G} = \frac{\Delta\tau \cdot \Delta U_{\rm p}}{\hbar}, \tag{24}$$

and the spontaneous-Raman-photon generation rate is derived as follows:

$$\frac{\Delta n_{\rm G}}{\Delta\tau} = \frac{\Delta U_{\rm p}}{\hbar} = -\frac{2}{\hbar}\varepsilon_o{\rm Im}\left\{\chi_{\rm SRG}^{(3)}\right\}|E_{\rm i}|^2|V_{\rm s}|^2 d\Omega = -\frac{2}{\hbar}\varepsilon_0{\rm Im}\left\{\chi_{\rm SRG}^{(3)}\right\}\frac{|\alpha_{\rm i}|^2\hbar\omega_{\rm i}}{2\varepsilon_o\mathcal{V}}\frac{\hbar\omega_{\rm s}}{2\varepsilon_o\mathcal{V}}d\Omega. \tag{25}$$

Figure 2 presents eight Feynman diagrams necessary for the calculations of the values of $\chi_{\rm SRL}^{(3)}$ and $\chi_{\rm SRG}^{(3)}$, related to the spontaneous Raman scattering with the transition from the initial state $|{\rm g}\rangle\langle{\rm g}|$. Surprisingly, in the Feynman diagrams of sRL, the vacuum field interacts with the target molecules twice; however, sRL indicates the coherent nature and travels in the same direction as the pump field. The eight terms in Fig. 2 occur simultaneously, and particularly, it is important to consider the upper term for sRL and lower term for sRG, as a pair. In nonresonant Raman scattering, the upper and lower terms in a pair indicate identical absolute values with the signs reversed. The term G1 in Fig. 2 represents a major contribution (rotating term: four arrows in energy level diagram lie above the ground state) to the sRG with the transition $|{\rm g}\rangle\langle{\rm g}| \rightarrow |{\rm v}\rangle\langle{\rm v}|$, which some references[5, 6] introduce. The term G2 expresses a minor contribution (half-rotating term: two arrows lie below the ground state) with the transition $|{\rm g}\rangle\langle{\rm g}| \rightarrow |{\rm g}\rangle\langle{\rm g}|$. The term G3 indicates another minor contribution (half-rotating term) with the transition $|{\rm g}\rangle\langle{\rm g}| \rightarrow |{\rm v}\rangle\langle{\rm v}|$, and the term G4 represents a negligible contribution (counter-rotating term: four arrows lie below the ground state) with the transition |g⟩⟨g| → |g⟩⟨g|. Similarly, the term L1 represents a major contribution to sRL with the transition |g⟩⟨g| → |g⟩⟨g|; the term L2 expresses a minor contribution with the transition |g⟩⟨g| → |v⟩⟨v|; the term L3 indicates another minor contribution with the transition |g⟩⟨g| → |g⟩⟨g|; and the term L4 represents a negligible contribution with the transition |g⟩⟨g| → |v⟩⟨v|. Note that the final states in the terms L1, G2, L3, and G4 are the ground state |g⟩⟨g|; however, these interactions inevitably occur with the other four terms.

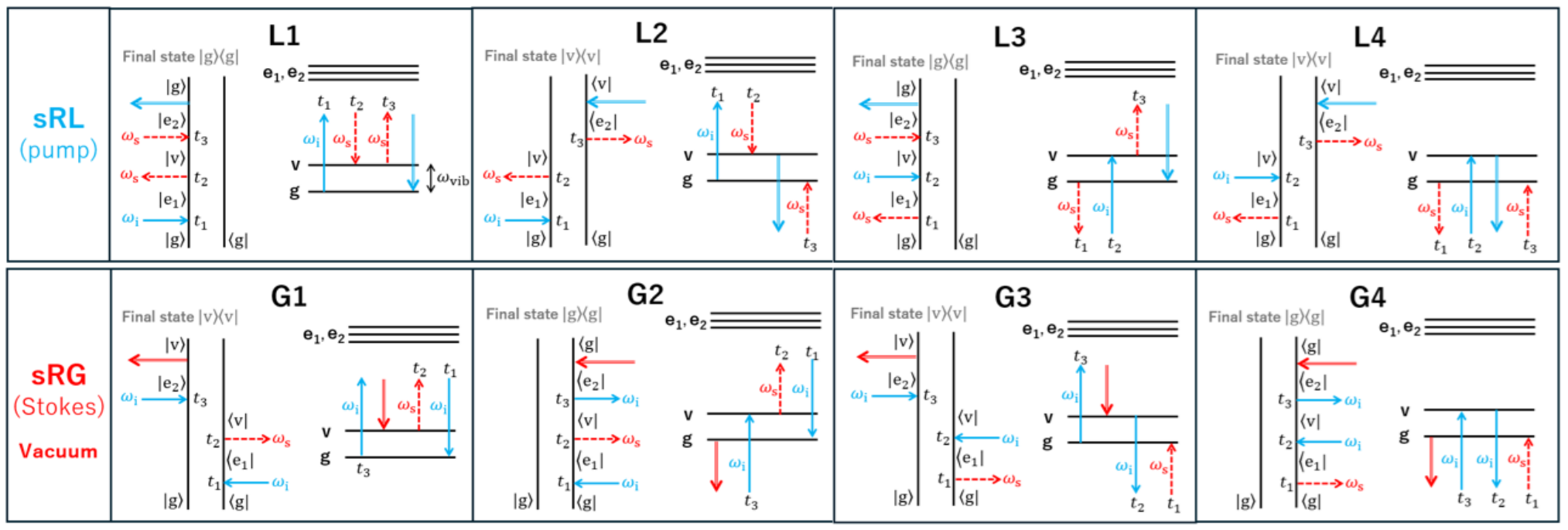

Fig. 2. Double-sided Feynman diagrams and energy level diagrams describing L1–L4 in sRL and G1–G4 in sRG. Dotted arrows mean the vacuum fields, solid arrows represent the pump fields, and double-line arrows denote the signal field. On the bra side (right side) in double-sided Feynman diagram, the arrows indicate the negative-frequency light going in and out of the target molecules, which corresponds to the positive-frequency light going out and in. Although the signal field is depicted in our diagram style to show the final state of molecules, the arrow for the signal field is not necessary to calculate the nonlinear susceptibility. Also in the energy level diagram, the corresponding arrows are drawn from left to right in the following order: starting from the bottom of the left line in the Feynman diagram and going up to the top, then moving to the right line and going down from top to bottom. In this style, the arrows on the energy level diagram can be drawn in a single stroke from left to right.

In each pair in Fig. 2, a photon disappears from the pump field through the sRL, a Raman-scattering photon emerges from the vacuum field through the sRG, and a molecular-vibration quanta remains through either sRL or sRG; i.e., each term in $\hbar\omega_{\mathrm{i}} = \hbar\omega_{\mathrm{s}} + \hbar\omega_{\mathrm{vib}}$ can be explained. Nonlinear optics provides the calculation rules [12 – 14] of the nonlinear susceptibility for the eight diagrams. Here a scalar theory is applied for comparison with the KHD formula. The terms L1–L4 in $\chi^{(3)}_{\mathrm{SRL}}$ and G1–G4 in $\chi^{(3)}_{\mathrm{SRG}}$ can be calculated as follows.

$$\chi^{(3)}_{\mathrm{L1}} = \frac{N}{\varepsilon_0\hbar^3}\sum_{\mathrm{e_1,e_2}} \frac{\mu_{\mathrm{ge_1}}\mu_{\mathrm{e_1v}}\mu_{\mathrm{ve_2}}\mu_{\mathrm{e_2g}}}{(\omega_{\mathrm{e_1g}} - \omega_{\mathrm{i}} - i\Gamma_{\mathrm{eg}})(\omega_{\mathrm{vib}} - \omega_{\mathrm{i}} + \omega_{\mathrm{s}} - i\gamma)(\omega_{\mathrm{e_2g}} - \omega_{\mathrm{i}} - i\Gamma_{\mathrm{eg}})} \tag{26}$$

$$\chi^{(3)}_{\mathrm{L2}} = \frac{N}{\varepsilon_0\hbar^3}\sum_{\mathrm{e_1,e_2}} \frac{(-1)\mu_{\mathrm{ge_1}}\mu_{\mathrm{e_1v}}\mu_{\mathrm{ve_2}}\mu_{\mathrm{e_2g}}}{(\omega_{\mathrm{e_1g}} - \omega_{\mathrm{i}} - i\Gamma_{\mathrm{eg}})(\omega_{\mathrm{vib}} - \omega_{\mathrm{i}} + \omega_{\mathrm{s}} - i\gamma)(-\omega_{\mathrm{e_2v}} - \omega_{\mathrm{i}} - i\Gamma_{\mathrm{ev}})} \tag{27}$$

$$\chi^{(3)}_{\mathrm{L3}} = \frac{N}{\varepsilon_0\hbar^3}\sum_{\mathrm{e_1,e_2}} \frac{\mu_{\mathrm{ge_1}}\mu_{\mathrm{e_1v}}\mu_{\mathrm{ve_2}}\mu_{\mathrm{e_2g}}}{(\omega_{\mathrm{e_1g}} + \omega_{\mathrm{s}} - i\Gamma_{\mathrm{eg}})(\omega_{\mathrm{vib}} + \omega_{\mathrm{s}} - \omega_{\mathrm{i}} - i\gamma)(\omega_{\mathrm{e_2g}} - \omega_{\mathrm{i}} - i\Gamma_{\mathrm{eg}})} \tag{28}$$

$$\chi^{(3)}_{\mathrm{L4}} = \frac{N}{\varepsilon_0\hbar^3}\sum_{\mathrm{e_1,e_2}} \frac{(-1)\mu_{\mathrm{ge_1}}\mu_{\mathrm{e_1v}}\mu_{\mathrm{ve_2}}\mu_{\mathrm{e_2g}}}{(\omega_{\mathrm{e_1g}} + \omega_{\mathrm{s}} - i\Gamma_{\mathrm{eg}})(\omega_{\mathrm{vib}} + \omega_{\mathrm{s}} - \omega_{\mathrm{i}} - i\gamma)(-\omega_{\mathrm{e_2v}} - \omega_{\mathrm{i}} - i\Gamma_{\mathrm{ev}})} \tag{29}$$

$$\chi^{(3)}_{\mathrm{G1}} = \frac{N}{\varepsilon_0\hbar^3}\sum_{\mathrm{e_1,e_2}} \frac{(-1)^2\mu_{\mathrm{ge_2}}\mu_{\mathrm{e_2v}}\mu_{\mathrm{ve_1}}\mu_{\mathrm{e_1g}}}{(-\omega_{\mathrm{e_1g}} + \omega_{\mathrm{i}} - i\Gamma_{\mathrm{eg}})(-\omega_{\mathrm{vib}} + \omega_{\mathrm{i}} - \omega_{\mathrm{s}} - i\gamma)(\omega_{\mathrm{e_2v}} - \omega_{\mathrm{s}} - i\Gamma_{\mathrm{ev}})} \tag{30}$$

$$\chi_{\mathrm{G2}}^{(3)} = \frac{N}{\varepsilon_0 \hbar^3} \sum_{\mathrm{e_1,e_2}} \frac{(-1)^3 \mu_{\mathrm{ge_2}} \mu_{\mathrm{e_2v}} \mu_{\mathrm{ve_1}} \mu_{\mathrm{e_1g}}}{(-\omega_{\mathrm{e_1g}} + \omega_{\mathrm{i}} - i\Gamma_{\mathrm{eg}})(-\omega_{\mathrm{vib}} + \omega_{\mathrm{i}} - \omega_{\mathrm{s}} - i\gamma)(-\omega_{\mathrm{e_2g}} - \omega_{\mathrm{s}} - i\Gamma_{\mathrm{eg}})} \quad (31)$$

$$\chi_{\mathrm{G3}}^{(3)} = \frac{N}{\varepsilon_0 \hbar^3} \sum_{\mathrm{e_1,e_2}} \frac{(-1)^2 \mu_{\mathrm{ge_2}} \mu_{\mathrm{e_2v}} \mu_{\mathrm{ve_1}} \mu_{\mathrm{e_1g}}}{(-\omega_{\mathrm{e_1g}} - \omega_{\mathrm{s}} - i\Gamma_{\mathrm{eg}})(-\omega_{\mathrm{vib}} - \omega_{\mathrm{s}} + \omega_{\mathrm{i}} - i\gamma)(\omega_{\mathrm{e_2v}} - \omega_{\mathrm{s}} - i\Gamma_{\mathrm{ev}})} \quad (32)$$

$$\chi_{\mathrm{G4}}^{(3)} = \frac{N}{\varepsilon_0 \hbar^3} \sum_{\mathrm{e_1,e_2}} \frac{(-1)^3 \mu_{\mathrm{ge_2}} \mu_{\mathrm{e_2v}} \mu_{\mathrm{ve_1}} \mu_{\mathrm{e_1g}}}{(-\omega_{\mathrm{e_1g}} - \omega_{\mathrm{s}} - i\Gamma_{\mathrm{eg}})(-\omega_{\mathrm{vib}} - \omega_{\mathrm{s}} + \omega_{\mathrm{i}} - i\gamma)(-\omega_{\mathrm{e_2g}} - \omega_{\mathrm{s}} - i\Gamma_{\mathrm{eg}})} \quad (33)$$

Here, $\Gamma_{\mathrm{eg}}$ is a damping factor for the polarized state between the states $|\mathrm{e}\rangle$ and $|\mathrm{g}\rangle$; $\Gamma_{\mathrm{ev}}$ is a damping factor for the polarized state between the states $|\mathrm{e}\rangle$ and $|\mathrm{v}\rangle$; $\mu_{\mathrm{ba}} = \langle \mathrm{b}|\mu|\mathrm{a}\rangle$ represents a transition dipole moment from the energy level a to the level b. The electronically excited levels $\mathrm{e_1}$ and $\mathrm{e_2}$ are indistinguishable from each other. The imaginary part of $\chi_{\mathrm{G1}}^{(3)}$ becomes

$$\begin{aligned}
-\mathrm{Im}\left\{\chi_{\mathrm{G1}}^{(3)}\right\} &= -\frac{N}{\varepsilon_0 \hbar^3} \mathrm{Im}\left\{\frac{1}{\omega_{\mathrm{vib}} - \omega_{\mathrm{i}} + \omega_{\mathrm{s}} + i\gamma}\right\} \\
&\quad \times \sum_{\mathrm{e_1,e_2}} \frac{\mu_{\mathrm{ge_2}} \mu_{\mathrm{e_2v}} \mu_{\mathrm{ve_1}} \mu_{\mathrm{e_1g}} \left\{(\omega_{\mathrm{e_2v}} - \omega_{\mathrm{s}})(\omega_{\mathrm{e_1g}} - \omega_{\mathrm{i}}) + \Gamma_{\mathrm{ev}}\Gamma_{\mathrm{eg}}\right\}}{\left\{(\omega_{\mathrm{e_2v}} - \omega_{\mathrm{s}})^2 + \Gamma_{\mathrm{ev}}^2\right\}\left\{(\omega_{\mathrm{e_1g}} - \omega_{\mathrm{i}})^2 + \Gamma_{\mathrm{eg}}^2\right\}} \\
&= \frac{-N}{\varepsilon_0 \hbar^3} \frac{-\gamma}{(\omega_{\mathrm{vib}} - \omega_{\mathrm{i}} + \omega_{\mathrm{s}})^2 + \gamma^2} \\
&\quad \times \sum_{\mathrm{e_1,e_2}} \frac{\mu_{\mathrm{ge_2}} \mu_{\mathrm{e_2v}} \mu_{\mathrm{ve_1}} \mu_{\mathrm{e_1g}} \left\{(\omega_{\mathrm{e_2v}} - \omega_{\mathrm{s}})(\omega_{\mathrm{e_1g}} - \omega_{\mathrm{i}}) + \Gamma_{\mathrm{ev}}\Gamma_{\mathrm{eg}}\right\}}{\left\{(\omega_{\mathrm{e_2v}} - \omega_{\mathrm{s}})^2 + \Gamma_{\mathrm{ev}}^2\right\}\left\{(\omega_{\mathrm{e_1g}} - \omega_{\mathrm{i}})^2 + \Gamma_{\mathrm{eg}}^2\right\}} \\
&\to \frac{N}{\varepsilon_0 \hbar^3} \pi\delta(\omega_{\mathrm{vib}} - \omega_{\mathrm{i}} + \omega_{\mathrm{s}}) \\
&\quad \times \sum_{\mathrm{e_1,e_2}} \frac{\mu_{\mathrm{ge_2}} \mu_{\mathrm{e_2v}} \mu_{\mathrm{ve_1}} \mu_{\mathrm{e_1g}} \left\{(\omega_{\mathrm{e_2v}} - \omega_{\mathrm{s}})(\omega_{\mathrm{e_1g}} - \omega_{\mathrm{i}}) + \Gamma_{\mathrm{ev}}\Gamma_{\mathrm{eg}}\right\}}{\left\{(\omega_{\mathrm{e_2v}} - \omega_{\mathrm{s}})^2 + \Gamma_{\mathrm{ev}}^2\right\}\left\{(\omega_{\mathrm{e_1g}} - \omega_{\mathrm{i}})^2 + \Gamma_{\mathrm{eg}}^2\right\}} \quad (34)
\end{aligned}$$

Here, we assume $\gamma \to +0$, which denotes a damping factor of the polarized state $|\mathrm{g}\rangle\langle\mathrm{v}|$ or $|\mathrm{v}\rangle\langle\mathrm{g}|$, to compare KHD theory. Similarly, other terms in the sRG can be calculated as follows.

$$\begin{aligned}
-\mathrm{Im}\left\{\chi_{\mathrm{G2}}^{(3)}\right\} &= \frac{N}{\varepsilon_0 \hbar^3} \pi\delta(\omega_{\mathrm{vib}} - \omega_{\mathrm{i}} + \omega_{\mathrm{s}}) \\
&\quad \times \sum_{\mathrm{e_1,e_2}} \frac{\mu_{\mathrm{ge_2}} \mu_{\mathrm{e_2v}} \mu_{\mathrm{ve_1}} \mu_{\mathrm{e_1g}} \left\{(\omega_{\mathrm{e_2g}} + \omega_{\mathrm{s}})(\omega_{\mathrm{e_1g}} - \omega_{\mathrm{i}}) - \Gamma_{\mathrm{eg}}\Gamma_{\mathrm{eg}}\right\}}{\left\{(\omega_{\mathrm{e_2g}} + \omega_{\mathrm{s}})^2 + \Gamma_{\mathrm{eg}}^2\right\}\left\{(\omega_{\mathrm{e_1g}} - \omega_{\mathrm{i}})^2 + \Gamma_{\mathrm{eg}}^2\right\}} \quad (35)
\end{aligned}$$

$$\begin{aligned}
-\mathrm{Im}\left\{\chi_{\mathrm{G3}}^{(3)}\right\} &= \frac{N}{\varepsilon_0 \hbar^3} \pi\delta(\omega_{\mathrm{vib}} - \omega_{\mathrm{i}} + \omega_{\mathrm{s}}) \\
&\quad \times \sum_{\mathrm{e_1,e_2}} \frac{\mu_{\mathrm{ge_2}} \mu_{\mathrm{e_2v}} \mu_{\mathrm{ve_1}} \mu_{\mathrm{e_1g}} \left\{(\omega_{\mathrm{e_2v}} - \omega_{\mathrm{s}})(\omega_{\mathrm{e_1g}} + \omega_{\mathrm{s}}) + \Gamma_{\mathrm{ev}}\Gamma_{\mathrm{eg}}\right\}}{\left\{(\omega_{\mathrm{e_2v}} - \omega_{\mathrm{s}})^2 + \Gamma_{\mathrm{ev}}^2\right\}\left\{(\omega_{\mathrm{e_1g}} + \omega_{\mathrm{s}})^2 + \Gamma_{\mathrm{eg}}^2\right\}} \quad (36)
\end{aligned}$$

$$\begin{aligned}
-\mathrm{Im}\left\{\chi_{\mathrm{G4}}^{(3)}\right\} &= \frac{N}{\varepsilon_0 \hbar^3} \pi\delta(\omega_{\mathrm{vib}} - \omega_{\mathrm{i}} + \omega_{\mathrm{s}}) \\
&\quad \times \sum_{\mathrm{e_1,e_2}} \frac{\mu_{\mathrm{ge_2}} \mu_{\mathrm{e_2v}} \mu_{\mathrm{ve_1}} \mu_{\mathrm{e_1g}} \left\{(\omega_{\mathrm{e_2g}} + \omega_{\mathrm{s}})(\omega_{\mathrm{e_1g}} + \omega_{\mathrm{s}}) - \Gamma_{\mathrm{eg}}\Gamma_{\mathrm{eg}}\right\}}{\left\{(\omega_{\mathrm{e_2g}} + \omega_{\mathrm{s}})^2 + \Gamma_{\mathrm{eg}}^2\right\}\left\{(\omega_{\mathrm{e_1g}} + \omega_{\mathrm{s}})^2 + \Gamma_{\mathrm{eg}}^2\right\}} \quad (37)
\end{aligned}$$

The terms in sRL can be calculated as follows.

$$-\mathrm{Im}\left\{\chi_{\mathrm{L1}}^{(3)}\right\} = -\frac{N}{\varepsilon_0\hbar^3}\mathrm{Im}\left\{\frac{1}{\omega_{\mathrm{vib}}-\omega_{\mathrm{i}}+\omega_{\mathrm{s}}-i\gamma}\right\}$$
$$\times\sum_{\mathrm{e_1,e_2}}\frac{\mu_{\mathrm{ge_1}}\mu_{\mathrm{e_1v}}\mu_{\mathrm{ve_2}}\mu_{\mathrm{e_2g}}\left\{\left(\omega_{\mathrm{e_2g}}-\omega_{\mathrm{i}}\right)\left(\omega_{\mathrm{e_1g}}-\omega_{\mathrm{i}}\right)-\Gamma_{\mathrm{eg}}\Gamma_{\mathrm{eg}}\right\}}{\left\{\left(\omega_{\mathrm{e_2g}}-\omega_{\mathrm{i}}\right)^2+\Gamma_{\mathrm{eg}}^2\right\}\left\{\left(\omega_{\mathrm{e_1g}}-\omega_{\mathrm{i}}\right)^2+\Gamma_{\mathrm{eg}}^2\right\}}$$
$$= \frac{-N}{\varepsilon_0\hbar^3}\pi\delta(\omega_{\mathrm{vib}}-\omega_{\mathrm{i}}+\omega_{\mathrm{s}})$$
$$\times\sum_{\mathrm{e_1,e_2}}\frac{\mu_{\mathrm{ge_1}}\mu_{\mathrm{e_1v}}\mu_{\mathrm{ve_2}}\mu_{\mathrm{e_2g}}\left\{\left(\omega_{\mathrm{e_2g}}-\omega_{\mathrm{i}}\right)\left(\omega_{\mathrm{e_1g}}-\omega_{\mathrm{i}}\right)-\Gamma_{\mathrm{eg}}\Gamma_{\mathrm{eg}}\right\}}{\left\{\left(\omega_{\mathrm{e_2g}}-\omega_{\mathrm{i}}\right)^2+\Gamma_{\mathrm{eg}}^2\right\}\left\{\left(\omega_{\mathrm{e_1g}}-\omega_{\mathrm{i}}\right)^2+\Gamma_{\mathrm{eg}}^2\right\}} \tag{38}$$

$$-\mathrm{Im}\left\{\chi_{\mathrm{L2}}^{(3)}\right\} = \frac{-N}{\varepsilon_0\hbar^3}\pi\delta(\omega_{\mathrm{vib}}-\omega_{\mathrm{i}}+\omega_{\mathrm{s}})$$
$$\times\sum_{\mathrm{e_1,e_2}}\frac{\mu_{\mathrm{ge_1}}\mu_{\mathrm{e_1v}}\mu_{\mathrm{ve_2}}\mu_{\mathrm{e_2g}}\left\{\left(\omega_{\mathrm{e_2v}}+\omega_{\mathrm{i}}\right)\left(\omega_{\mathrm{e_1g}}-\omega_{\mathrm{i}}\right)+\Gamma_{\mathrm{ev}}\Gamma_{\mathrm{eg}}\right\}}{\left\{\left(\omega_{\mathrm{e_2v}}+\omega_{\mathrm{i}}\right)^2+\Gamma_{\mathrm{ev}}^2\right\}\left\{\left(\omega_{\mathrm{e_1g}}-\omega_{\mathrm{i}}\right)^2+\Gamma_{\mathrm{eg}}^2\right\}} \tag{39}$$

$$-\mathrm{Im}\left\{\chi_{\mathrm{L3}}^{(3)}\right\} = \frac{-N}{\varepsilon_0\hbar^3}\pi\delta(\omega_{\mathrm{vib}}-\omega_{\mathrm{i}}+\omega_{\mathrm{s}})$$
$$\times\sum_{\mathrm{e_1,e_2}}\frac{\mu_{\mathrm{ge_1}}\mu_{\mathrm{e_1v}}\mu_{\mathrm{ve_2}}\mu_{\mathrm{e_2g}}\left\{\left(\omega_{\mathrm{e_2g}}-\omega_{\mathrm{i}}\right)\left(\omega_{\mathrm{e_1g}}+\omega_{\mathrm{s}}\right)-\Gamma_{\mathrm{eg}}\Gamma_{\mathrm{eg}}\right\}}{\left\{\left(\omega_{\mathrm{e_2g}}-\omega_{\mathrm{i}}\right)^2+\Gamma_{\mathrm{eg}}^2\right\}\left\{\left(\omega_{\mathrm{e_1g}}+\omega_{\mathrm{s}}\right)^2+\Gamma_{\mathrm{eg}}^2\right\}} \tag{40}$$

$$-\mathrm{Im}\left\{\chi_{\mathrm{L4}}^{(3)}\right\} = \frac{-N}{\varepsilon_0\hbar^3}\pi\delta(\omega_{\mathrm{vib}}-\omega_{\mathrm{i}}+\omega_{\mathrm{s}})$$
$$\times\sum_{\mathrm{e_1,e_2}}\frac{\mu_{\mathrm{ge_1}}\mu_{\mathrm{e_1v}}\mu_{\mathrm{ve_2}}\mu_{\mathrm{e_2g}}\left\{\left(\omega_{\mathrm{e_2v}}+\omega_{\mathrm{i}}\right)\left(\omega_{\mathrm{e_1g}}+\omega_{\mathrm{s}}\right)+\Gamma_{\mathrm{ev}}\Gamma_{\mathrm{eg}}\right\}}{\left\{\left(\omega_{\mathrm{e_2v}}+\omega_{\mathrm{i}}\right)^2+\Gamma_{\mathrm{ev}}^2\right\}\left\{\left(\omega_{\mathrm{e_1g}}+\omega_{\mathrm{s}}\right)^2+\Gamma_{\mathrm{eg}}^2\right\}} \tag{41}$$

Considering the relations $\omega_{\mathrm{ev}}-\omega_{\mathrm{s}}=\omega_{\mathrm{eg}}-\omega_{\mathrm{i}}$ and $\omega_{\mathrm{ev}}+\omega_{\mathrm{i}}=\omega_{\mathrm{eg}}+\omega_{\mathrm{s}}$, the terms for G1 and L1 under the nonresonant condition become approximately equal, except for the sign. In an equivalent manner, the terms for G2, G3, L2, and L3 indicate approximately equal absolute values, except for the sign; the terms for G4 and L4 are also approximately equal, except for the sign.

To calculate the spontaneous-Raman-photon generation rate $\Delta n_{\mathrm{G}}/\Delta\tau$, we substitute

$$-\mathrm{Im}\left\{\chi_{\mathrm{SRG}}^{(3)}\right\} = -\mathrm{Im}\left\{\chi_{\mathrm{G1}}^{(3)}\right\}-\mathrm{Im}\left\{\chi_{\mathrm{G2}}^{(3)}\right\}-\mathrm{Im}\left\{\chi_{\mathrm{G3}}^{(3)}\right\}-\mathrm{Im}\left\{\chi_{\mathrm{G4}}^{(3)}\right\} \tag{42}$$

into Eq. (25) and obtain

$$\frac{\Delta n_{\mathrm{G}}}{\Delta\tau} = -\frac{2}{\hbar}\varepsilon_0\mathrm{Im}\left\{\chi_{\mathrm{SRG}}^{(3)}\right\}\frac{|\alpha_{\mathrm{i}}|^2\hbar\omega_{\mathrm{i}}}{2\varepsilon_{\mathrm{o}}\mathcal{V}}\frac{\hbar\omega_{\mathrm{s}}}{2\varepsilon_{\mathrm{o}}\mathcal{V}}d\Omega$$
$$= G\frac{2}{\hbar}\frac{N}{\hbar^3}\pi\delta(\omega_{\mathrm{vib}}-\omega_{\mathrm{i}}+\omega_{\mathrm{s}})\frac{|\alpha_{\mathrm{i}}|^2\hbar\omega_{\mathrm{i}}}{2\varepsilon_0\mathcal{V}}\frac{\hbar\omega_{\mathrm{s}}}{2\varepsilon_0\mathcal{V}}d\Omega, \tag{43}$$

with

$$G = g_1+g_2+g_3+g_4 \tag{44a}$$

$$g_1 = \sum_{\mathrm{e_1,e_2}}\frac{\mu_{\mathrm{ge_2}}\mu_{\mathrm{e_2v}}\mu_{\mathrm{ve_1}}\mu_{\mathrm{e_1g}}\left\{\left(\omega_{\mathrm{e_2g}}-\omega_{\mathrm{i}}\right)\left(\omega_{\mathrm{e_1g}}-\omega_{\mathrm{i}}\right)+\Gamma_{\mathrm{ev}}\Gamma_{\mathrm{eg}}\right\}}{\left\{\left(\omega_{\mathrm{e_2g}}-\omega_{\mathrm{i}}\right)^2+\Gamma_{\mathrm{ev}}^2\right\}\left\{\left(\omega_{\mathrm{e_1g}}-\omega_{\mathrm{i}}\right)^2+\Gamma_{\mathrm{eg}}^2\right\}} \tag{44b}$$

$$g_2 = \sum_{\mathrm{e_1,e_2}}\frac{\mu_{\mathrm{ge_2}}\mu_{\mathrm{e_2v}}\mu_{\mathrm{ve_1}}\mu_{\mathrm{e_1g}}\left\{\left(\omega_{\mathrm{e_2v}}+\omega_{\mathrm{i}}\right)\left(\omega_{\mathrm{e_1g}}-\omega_{\mathrm{i}}\right)-\Gamma_{\mathrm{eg}}\Gamma_{\mathrm{eg}}\right\}}{\left\{\left(\omega_{\mathrm{e_2v}}+\omega_{\mathrm{i}}\right)^2+\Gamma_{\mathrm{eg}}^2\right\}\left\{\left(\omega_{\mathrm{e_1g}}-\omega_{\mathrm{i}}\right)^2+\Gamma_{\mathrm{eg}}^2\right\}} \tag{44c}$$

$$g_3 = \sum_{e_1,e_2} \frac{\mu_{ge_2}\mu_{e_2v}\mu_{ve_1}\mu_{e_1g}\{(\omega_{e_2g} - \omega_i)(\omega_{e_1v} + \omega_i) + \Gamma_{ev}\Gamma_{eg}\}}{\{(\omega_{e_2g} - \omega_i)^2 + \Gamma_{ev}^2\}\{(\omega_{e_1v} + \omega_i)^2 + \Gamma_{eg}^2\}} \tag{44d}$$

$$g_4 = \sum_{e_1,e_2} \frac{\mu_{ge_2}\mu_{e_2v}\mu_{ve_1}\mu_{e_1g}\{(\omega_{e_2v} + \omega_i)(\omega_{e_1v} + \omega_i) - \Gamma_{eg}\Gamma_{eg}\}}{\{(\omega_{e_2v} + \omega_i)^2 + \Gamma_{eg}^2\}\{(\omega_{e_1v} + \omega_i)^2 + \Gamma_{eg}^2\}}. \tag{44e}$$

Here, we use the relations $\omega_{ev} - \omega_s = \omega_{eg} - \omega_i$ and $\omega_{ev} + \omega_i = \omega_{eg} + \omega_s$.

To transform the photon generation rate $\Delta n_G/\Delta\tau$ into the scattering cross section $\sigma_{\boldsymbol{k}_s}$, we normalize it with the expectation value of the photon numbers in the pump beam that flows through a unit area in a unit time $|\alpha_i|^2 c/\mathcal{V}$:

$$\begin{aligned}\sigma_{\boldsymbol{k}_s} &= \frac{\Delta n_G}{\Delta\tau}\frac{\mathcal{V}}{|\alpha_i|^2 c} \\ &= \int G \frac{2}{\hbar}\frac{N}{\hbar^3}\pi\delta(\omega_{vib} - \omega_i + \omega_s)\frac{|\alpha_i|^2\hbar\omega_i}{2\varepsilon_0\mathcal{V}}\frac{\hbar\omega_s}{2\varepsilon_0\mathcal{V}}\frac{\mathcal{V}}{|\alpha_i|^2 c}\,d\Omega.\end{aligned} \tag{45}$$

The differential cross section is expressed as

$$\frac{d\sigma_{\boldsymbol{k}_s}}{d\Omega} = G\frac{2N}{\hbar^4}\pi\delta(\omega_{vib} - \omega_i + \omega_s)\frac{|\alpha_i|^2\hbar\omega_i}{2\varepsilon_o\mathcal{V}}\frac{\hbar\omega_s}{2\varepsilon_o\mathcal{V}}\frac{\mathcal{V}}{|\alpha_i|^2 c}. \tag{46}$$

Note that the above equation represents the contribution from a single vacuum mode. Considering all vacuum-field modes $\boldsymbol{k}_s$, the differential cross section becomes

$$\begin{aligned}\frac{d\sigma_{sRG}}{d\Omega} &= \sum_{\boldsymbol{k}_s}\frac{d\sigma_{\boldsymbol{k}_s}}{d\Omega} \\ &= \sum_{\boldsymbol{k}_s} G\frac{2N}{\hbar^4}\pi\delta(\omega_{vib} - \omega_i + \omega_s)\frac{|\alpha_i|^2\hbar\omega_i}{2\varepsilon_o\mathcal{V}}\frac{\hbar\omega_s}{2\varepsilon_o\mathcal{V}}\frac{\mathcal{V}}{|\alpha_i|^2 c}.\end{aligned} \tag{47}$$

If we consider the vacuum-field mode density, the summation of $\boldsymbol{k}_s$ can be replaced by the integral of $\omega_s$ as follows [9,10].

$$\sum_{\boldsymbol{k}_s}(\cdots) \to \int(\cdots)\left(\frac{\omega_s^2}{c^3}\times\frac{\mathcal{V}}{(2\pi)^3}\right)d\omega_s \tag{48}$$

Accordingly, the differential cross section of spontaneous Raman scattering can be obtained as

$$\frac{d\sigma_{sRG}}{d\Omega} = G\frac{N}{16\pi^2\varepsilon_o{}^2c^4}\frac{\omega_i(\omega_i - \omega_{vib})^3}{\hbar^2}. \tag{49}$$

Likewise, the differential cross section of spontaneous Raman loss can be written as

$$\frac{d\sigma_{sRL}}{d\Omega} = L\frac{N}{16\pi^2\varepsilon_o{}^2c^4}\frac{\omega_i(\omega_i - \omega_{vib})^3}{\hbar^2}, \tag{50}$$

with

$$L = l_1 + l_2 + l_3 + l_4 \tag{51a}$$

$$l_1 = \sum_{e_1,e_2}\frac{\mu_{ge_1}\mu_{e_1v}\mu_{ve_2}\mu_{e_2g}\{(\omega_{e_2g} - \omega_i)(\omega_{e_1g} - \omega_i) - \Gamma_{eg}\Gamma_{eg}\}}{\{(\omega_{e_2g} - \omega_i)^2 + \Gamma_{eg}^2\}\{(\omega_{e_1g} - \omega_i)^2 + \Gamma_{eg}^2\}} \tag{51b}$$

$$l_2 = \sum_{e_1,e_2}\frac{\mu_{ge_1}\mu_{e_1v}\mu_{ve_2}\mu_{e_2g}\{(\omega_{e_2v} + \omega_i)(\omega_{e_1g} - \omega_i) + \Gamma_{ev}\Gamma_{eg}\}}{\{(\omega_{e_2v} + \omega_i)^2 + \Gamma_{ev}^2\}\{(\omega_{e_1g} - \omega_i)^2 + \Gamma_{eg}^2\}} \tag{51c}$$

$$l_3 = \sum_{e_1,e_2}\frac{\mu_{ge_1}\mu_{e_1v}\mu_{ve_2}\mu_{e_2g}\{(\omega_{e_2g} - \omega_i)(\omega_{e_1v} + \omega_i) - \Gamma_{eg}\Gamma_{eg}\}}{\{(\omega_{e_2g} - \omega_i)^2 + \Gamma_{eg}^2\}\{(\omega_{e_1v} + \omega_i)^2 + \Gamma_{eg}^2\}} \tag{51d}$$

$$l_4 = \sum_{e_1,e_2} \frac{\mu_{ge_1}\mu_{e_1v}\mu_{ve_2}\mu_{e_2g}\left\{\left(\omega_{e_2v}+\omega_i\right)\left(\omega_{e_1v}+\omega_i\right)+\Gamma_{ev}\Gamma_{eg}\right\}}{\left\{\left(\omega_{e_2v}+\omega_i\right)^2+\Gamma_{ev}^2\right\}\left\{\left(\omega_{e_1v}+\omega_i\right)^2+\Gamma_{eg}^2\right\}}. \tag{51e}$$

Here, according to the definition of the cross section, the sign is changed to positive.

# III. DISCUSSION

## A. Comparison with traditional KHD theory and nonlinear optical approach

In KHD theory, considering that $T_1$ and $T_4$ are real numbers and $T_2 = T_3^*$, we obtain

$$\begin{aligned}\frac{d\sigma_{\mathrm{KHD}}}{d\Omega} &= A(T_1 + \mathrm{Re}\{T_2\} + \mathrm{Re}\{T_3\} + T_4)\\ &= A(\mathrm{Re}\{T_1\} + \mathrm{Re}\{T_2\} + \mathrm{Re}\{T_3\} + \mathrm{Re}\{T_4\}),\end{aligned} \tag{52}$$

with

$$A = \frac{N}{16\pi^2{\varepsilon_0}^2c^4}\frac{\omega_i(\omega_i-\omega_{vib})^3}{\hbar^2} \tag{53a}$$

$$\mathrm{Re}\{T_1\} = A\sum_{e_1,e_2} \frac{\mu_{ge_2}\mu_{e_2v}\mu_{ve_1}\mu_{e_1g}\left\{\left(\omega_{e_2g}-\omega_i\right)\left(\omega_{e_1g}-\omega_i\right)+\Gamma^2\right\}}{\left\{\left(\omega_{e_2g}-\omega_i\right)^2+\Gamma^2\right\}\left\{\left(\omega_{e_1g}-\omega_i\right)^2+\Gamma^2\right\}} \tag{53b}$$

$$\mathrm{Re}\{T_2\} = A\sum_{e_1,e_2} \frac{\mu_{ge_2}\mu_{e_2v}\mu_{ve_1}\mu_{e_1g}\left\{\left(\omega_{e_2g}-\omega_i\right)\left(\omega_{e_1v}+\omega_i\right)+\Gamma^2\right\}}{\left\{\left(\omega_{e_2g}-\omega_i\right)^2+\Gamma^2\right\}\left\{\left(\omega_{e_1v}+\omega_i\right)^2+\Gamma^2\right\}} \tag{53c}$$

$$\mathrm{Re}\{T_3\} = A\sum_{e_1,e_2} \frac{\mu_{ge_1}\mu_{e_1v}\mu_{ve_2}\mu_{e_2g}\left\{\left(\omega_{e_2v}+\omega_i\right)\left(\omega_{e_1g}-\omega_i\right)+\Gamma^2\right\}}{\left\{\left(\omega_{e_2v}+\omega_i\right)^2+\Gamma^2\right\}\left\{\left(\omega_{e_1g}-\omega_i\right)^2+\Gamma^2\right\}} \tag{53d}$$

$$\mathrm{Re}\{T_4\} = A\sum_{e_1,e_2} \frac{\mu_{ge_1}\mu_{e_1v}\mu_{ve_2}\mu_{e_2g}\left\{\left(\omega_{e_2v}+\omega_i\right)\left(\omega_{e_1v}+\omega_i\right)+\Gamma^2\right\}}{\left\{\left(\omega_{e_2v}+\omega_i\right)^2+\Gamma^2\right\}\left\{\left(\omega_{e_1v}+\omega_i\right)^2+\Gamma^2\right\}}. \tag{53e}$$

Since the approximation $\gamma \to +0$ is implicitly used in KHD theory, a relation $\Gamma = \Gamma_{ev} = \Gamma_{eg}$ can be assumed. Using the notation defined below:

$$\begin{gathered} G1 = A\,g_1\\ G2 = A\,g_2\\ G3 = A\,g_3\\ G4 = A\,g_4\\ L1 = A\,l_1\\ L2 = A\,l_2\\ L3 = A\,l_3\\ L4 = A\,l_4,\end{gathered} \tag{54}$$

and comparing Eqs. (49), (50), and (52), the following relationship can be derived:

$$\begin{gathered} \mathrm{Re}\{T_1\} = G1\\ \mathrm{Re}\{T_2\} = L2\\ \mathrm{Re}\{T_3\} = G3\\ \mathrm{Re}\{T_4\} = L4,\end{gathered} \tag{55}$$

i.e.,

$$\frac{d\sigma_{\mathrm{KHD}}}{d\Omega} = G1 + L2 + G3 + L4. \tag{56}$$

For reference, the differential cross sections of sRG and sRL are expressed as

$$\frac{d\sigma_{\mathrm{sRG}}}{d\Omega} = G1 + G2 + G3 + G4 \tag{57}$$
$$\frac{d\sigma_{\mathrm{sRL}}}{d\Omega} = L1 + L2 + L3 + L4. \tag{58}$$

As discussed above, the optical processes sRL and sRG occur simultaneously, with the relations presented below.

$$-\mathrm{Im}\left\{\chi^{(3)}_{\mathrm{G1}}\right\} \approx \mathrm{Im}\left\{\chi^{(3)}_{\mathrm{L1}}\right\}$$
$$-\mathrm{Im}\left\{\chi^{(3)}_{\mathrm{G2}}\right\} \approx \mathrm{Im}\left\{\chi^{(3)}_{\mathrm{L2}}\right\}$$
$$-\mathrm{Im}\left\{\chi^{(3)}_{\mathrm{G3}}\right\} \approx \mathrm{Im}\left\{\chi^{(3)}_{\mathrm{L3}}\right\}$$
$$-\mathrm{Im}\left\{\chi^{(3)}_{\mathrm{G4}}\right\} \approx \mathrm{Im}\left\{\chi^{(3)}_{\mathrm{L4}}\right\} \tag{59}$$

It should be noted that these relations are only for nonresonant Raman scattering, where the damping factors $\Gamma_{\mathrm{ev}}$ and $\Gamma_{\mathrm{eg}}$ can be ignored, and that the differential cross sections for the sRL must be defined considering the sign opposite to that of sRG such that the cross section becomes positive.

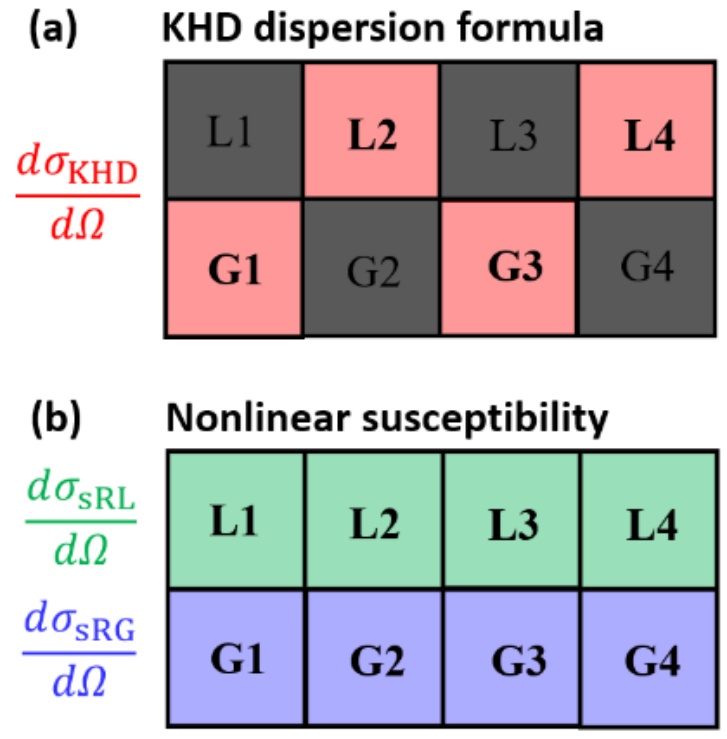

Fig. 3. A relationship between the cross sections. (a) $\sigma_{\mathrm{KHD}}/\Omega$ derived from KHD dispersion formula. (b) $\sigma_{\mathrm{sRG}}/\Omega$ and $\sigma_{\mathrm{sRL}}/\Omega$ derived from nonlinear optical susceptibility approach.

It appears that the susceptibility for spontaneous Raman scattering that corresponds to the KHD dispersion formula is

$$-\mathrm{Im}\left\{\chi^{(3)}_{\mathrm{G1}}\right\} + \mathrm{Im}\left\{\chi^{(3)}_{\mathrm{L2}}\right\} - \mathrm{Im}\left\{\chi^{(3)}_{\mathrm{G3}}\right\} + \mathrm{Im}\left\{\chi^{(3)}_{\mathrm{L4}}\right\}. \tag{60}$$

In the processes corresponding to G1, G2, L3, and L4, the system undergoes a transition from the initial state $|\mathrm{g}\rangle\langle\mathrm{g}|$ to the final state $|\mathrm{v}\rangle\langle\mathrm{v}|$, leaving molecular vibrational quanta in the system. Conversely, the interactions L1, L2, G3, and G4 result in the final state $|\mathrm{g}\rangle\langle\mathrm{g}|$, leaving no residual vibrational quanta. From this perspective, the KHD framework naturally defines the probability of vibrational quantum generation through Eq. (1) [or equivalently, Eq. (52) or Eq. (56)]. Consequently, the KHD dispersion formulation is limited to describing this generation probability, leaving the pump-photon depletion rate and the Stokes-photon generation rate unaddressed.

If an experimental setup approaches the electronical-resonance condition, the deferential cross section changes. Considering that $\Gamma_{\mathrm{ev}}$ and $\Gamma_{\mathrm{eg}}$ have small finite values and that they do not approach zero but $\gamma$ approaches zero, i.e., $\gamma \to +0$, and defining $\epsilon = \omega_{\mathrm{vib}} - \omega_{\mathrm{i}} + \omega_{\mathrm{s}} = 0$, the nonlinear susceptibility for G1 in sRG becomes

$$
\begin{aligned}
&-\mathrm{Im}\left\{\chi_{\mathrm{G1}}^{(3)}\right\} \\
&= -\frac{N}{\varepsilon_0\hbar^3}\mathrm{Im}\left\{\frac{\mu_{\mathrm{ge_2}}\mu_{\mathrm{e_2v}}\mu_{\mathrm{ve_1}}\mu_{\mathrm{e_1g}}}{(\omega_{\mathrm{vib}}-\omega_{\mathrm{i}}+\omega_{\mathrm{s}}+i\gamma)(\omega_{\mathrm{e_2v}}-\omega_{\mathrm{s}}-i\Gamma_{\mathrm{ev}})(\omega_{\mathrm{e_1g}}-\omega_{\mathrm{i}}+i\Gamma_{\mathrm{eg}})}\right\} \\
&= -\frac{N}{\varepsilon_0\hbar^3}\mathrm{Im}\left\{\left(\frac{\epsilon}{\epsilon^2+\gamma^2}-i\frac{\gamma}{\epsilon^2+\gamma^2}\right)\frac{\mu_{\mathrm{ge_2}}\mu_{\mathrm{e_2v}}\mu_{\mathrm{ve_1}}\mu_{\mathrm{e_1g}}}{(-i\Gamma_{\mathrm{ev}})(+i\Gamma_{\mathrm{eg}})}\right\} \\
&= -\frac{N}{\varepsilon_0\hbar^3}\mathrm{Im}\left\{\left(\frac{\epsilon}{\gamma^2}-i\pi\delta(\epsilon)\right)\frac{\mu_{\mathrm{ge_2}}\mu_{\mathrm{e_2v}}\mu_{\mathrm{ve_1}}\mu_{\mathrm{e_1g}}}{(-i\Gamma_{\mathrm{ev}})(+i\Gamma_{\mathrm{eg}})}\right\} \\
&= \frac{N}{\varepsilon_0\hbar^3}\pi\delta(\omega_{\mathrm{vib}}-\omega_{\mathrm{i}}+\omega_{\mathrm{s}})\frac{\mu_{\mathrm{ge_2}}\mu_{\mathrm{e_2v}}\mu_{\mathrm{ve_1}}\mu_{\mathrm{e_1g}}}{\Gamma_{\mathrm{ev}}\,\Gamma_{\mathrm{eg}}} \qquad (61)
\end{aligned}
$$

where we used the resonance condition $\omega_{\mathrm{e_2v}}-\omega_{\mathrm{s}}=0$ and $\omega_{\mathrm{e_1g}}-\omega_{\mathrm{i}}=0$. Likewise, the nonlinear susceptibility for L1 in sRL can be calculated as follows.

$$
\begin{aligned}
&-\mathrm{Im}\left\{\chi_{\mathrm{L1}}^{(3)}\right\} \\
&= -\frac{N}{\varepsilon_0\hbar^3}\mathrm{Im}\left\{\frac{\mu_{\mathrm{ge_1}}\mu_{\mathrm{e_1v}}\mu_{\mathrm{ve_2}}\mu_{\mathrm{e_2g}}}{(\omega_{\mathrm{vib}}-\omega_{\mathrm{i}}+\omega_{\mathrm{s}}-i\gamma)(\omega_{\mathrm{e_2g}}-\omega_{\mathrm{i}}-i\Gamma_{\mathrm{eg}})(\omega_{\mathrm{e_1g}}-\omega_{\mathrm{i}}-i\Gamma_{\mathrm{eg}})}\right\} \\
&= -\frac{N}{\varepsilon_0\hbar^3}\mathrm{Im}\left\{\left(\frac{\epsilon}{\epsilon^2+\gamma^2}+i\frac{\gamma}{\epsilon^2+\gamma^2}\right)\frac{\mu_{\mathrm{ge_1}}\mu_{\mathrm{e_1v}}\mu_{\mathrm{ve_2}}\mu_{\mathrm{e_2g}}}{(-i\Gamma_{\mathrm{eg}})(-i\Gamma_{\mathrm{eg}})}\right\} \\
&= -\frac{N}{\varepsilon_0\hbar^3}\mathrm{Im}\left\{\left(\frac{\epsilon}{\gamma^2}+i\pi\delta(\epsilon)\right)\frac{\mu_{\mathrm{ge_1}}\mu_{\mathrm{e_1v}}\mu_{\mathrm{ve_2}}\mu_{\mathrm{e_2g}}}{(-i\Gamma_{\mathrm{eg}})(-i\Gamma_{\mathrm{eg}})}\right\} \\
&= \frac{N}{\varepsilon_0\hbar^3}\pi\delta(\omega_{\mathrm{vib}}-\omega_{\mathrm{i}}+\omega_{\mathrm{s}})\frac{\mu_{\mathrm{ge_1}}\mu_{\mathrm{e_1v}}\mu_{\mathrm{ve_2}}\mu_{\mathrm{e_2g}}}{\Gamma_{\mathrm{eg}}\,\Gamma_{\mathrm{eg}}} \qquad (62)
\end{aligned}
$$

Under the resonance condition, the sRL turns to a gain process with the same wavelength as the pump beam, which plays a role as a saturation phenomenon for $\chi^{(1)}$ linear absorption. In an analogous manner, the imaginary part of the nonlinear susceptibility for G2 can be calculated as follows.

$$
\begin{aligned}
&-\mathrm{Im}\left\{\chi_{\mathrm{G3}}^{(3)}\right\} \\
&= -\frac{N}{\varepsilon_0\hbar^3}\mathrm{Im}\left\{\frac{\mu_{\mathrm{ge_2}}\mu_{\mathrm{e_2v}}\mu_{\mathrm{ve_1}}\mu_{\mathrm{e_1g}}}{(\omega_{\mathrm{vib}}-\omega_{\mathrm{i}}+\omega_{\mathrm{s}}+i\gamma)(\omega_{\mathrm{e_2v}}-\omega_{\mathrm{s}}-i\Gamma_{\mathrm{ev}})(\omega_{\mathrm{e_1g}}+\omega_{\mathrm{s}}+i\Gamma_{\mathrm{eg}})}\right\} \\
&= -\frac{N}{\varepsilon_0\hbar^3}\mathrm{Im}\left\{\left(\frac{\epsilon}{\epsilon^2+\gamma^2}-i\frac{\gamma}{\epsilon^2+\gamma^2}\right)\frac{\mu_{\mathrm{ge_2}}\mu_{\mathrm{e_2v}}\mu_{\mathrm{ve_1}}\mu_{\mathrm{e_1g}}}{(-i\Gamma_{\mathrm{ev}})(\omega_{\mathrm{e_1g}}+\omega_{\mathrm{s}})}\right\} \\
&= -\frac{N}{\varepsilon_0\hbar^3}\mathrm{Im}\left\{\left(\frac{\epsilon}{\gamma^2}-i\pi\delta(\epsilon)\right)\frac{\mu_{\mathrm{ge_2}}\mu_{\mathrm{e_2v}}\mu_{\mathrm{ve_1}}\mu_{\mathrm{e_1g}}}{(-i\Gamma_{\mathrm{ev}})(\omega_{\mathrm{e_1g}}+\omega_{\mathrm{s}})}\right\} \\
&= -\frac{N}{\varepsilon_0\hbar^3}\left(\frac{\epsilon}{\gamma^2}\right)\frac{\mu_{\mathrm{ge_2}}\mu_{\mathrm{e_2v}}\mu_{\mathrm{ve_1}}\mu_{\mathrm{e_1g}}}{\Gamma_{\mathrm{ev}}\,(\omega_{\mathrm{e_1g}}+\omega_{\mathrm{s}})} \\
&= 0 \qquad (63)
\end{aligned}
$$

Likewise, the imaginary parts of the nonlinear susceptibilities for G3, L2, and L3 become

$$
-\mathrm{Im}\left\{\chi_{\mathrm{G2}}^{(3)}\right\} = -\mathrm{Im}\left\{\chi_{\mathrm{L2}}^{(3)}\right\} = -\mathrm{Im}\left\{\chi_{\mathrm{L3}}^{(3)}\right\} = 0. \qquad (64)
$$

Because of the large energy-denominator, $-\mathrm{Im}\left\{\chi_{\mathrm{G4}}^{(3)}\right\}$ and $-\mathrm{Im}\left\{\chi_{\mathrm{L4}}^{(3)}\right\}$ under the resonant condition become smaller than that under the nonresonant condition. Thus, under the resonant condition, the cross sections for G2–4 and L2–4 are negligible and G1 dominates the spontaneous Raman scattering (sRG), and in the sRL, L1 turns to a gain process and increases the pump beam intensity. Importantly, $\chi^{(1)}$ linear absorption inevitably proceeds concurrently, maintaining overall energy conservation.

## B. Unobservable terms of susceptibility

To establish a unified framework encompassing both the susceptibility approach and the KHD formulation—the generalization of which is detailed in Subsection C—we begin by deriving the susceptibilites associated with the *unobservable* Raman processes, where the molecular system transitions from the initial state $|\mathrm{v}\rangle\langle\mathrm{v}|$ to the final state $|\mathrm{g}\rangle\langle\mathrm{g}|$. Based on the formalism of nonlinear optics[12-14], eight diagrams corresponding to unobservable Raman scattering (L5–L8 and G5–G8) are illustrated in Fig. 4. The third-order susceptibilities associated with terms L5–L8 for $\chi^{(3)}_{\mathrm{SRL}}$ and G5–G8 for $\chi^{(3)}_{\mathrm{SRG}}$ are calculated as follows:

$$\chi^{(3)}_{\mathrm{G5}} = \frac{N}{\varepsilon_0 \hbar^3} \sum_{\mathrm{e_1,e_2}} \frac{(-1)^2 \mu_{\mathrm{ve_2}} \mu_{\mathrm{e_2g}} \mu_{\mathrm{ge_1}} \mu_{\mathrm{e_1v}}}{(-\omega_{\mathrm{e_1v}} + \omega_{\mathrm{s}} - i\Gamma_{\mathrm{ev}})(\omega_{\mathrm{vib}} - \omega_{\mathrm{i}} + \omega_{\mathrm{s}} - i\gamma)(\omega_{\mathrm{e_2g}} - \omega_{\mathrm{i}} - i\Gamma_{\mathrm{eg}})} \tag{65}$$

$$\chi^{(3)}_{\mathrm{G6}} = \frac{N}{\varepsilon_0 \hbar^3} \sum_{\mathrm{e_1,e_2}} \frac{(-1)^3 \mu_{\mathrm{ve_2}} \mu_{\mathrm{e_2g}} \mu_{\mathrm{ge_1}} \mu_{\mathrm{e_1v}}}{(-\omega_{\mathrm{e_1v}} + \omega_{\mathrm{s}} - i\Gamma_{\mathrm{ev}})(\omega_{\mathrm{vib}} + \omega_{\mathrm{s}} - \omega_{\mathrm{i}} - i\gamma)(-\omega_{\mathrm{e_2v}} - \omega_{\mathrm{i}} - i\Gamma_{\mathrm{ev}})} \tag{66}$$

$$\chi^{(3)}_{\mathrm{G7}} = \frac{N}{\varepsilon_0 \hbar^3} \sum_{\mathrm{e_1,e_2}} \frac{(-1)^2 \mu_{\mathrm{ve_2}} \mu_{\mathrm{e_2g}} \mu_{\mathrm{ge_1}} \mu_{\mathrm{e_1v}}}{(-\omega_{\mathrm{e_1v}} - \omega_{\mathrm{i}} - i\Gamma_{\mathrm{ev}})(\omega_{\mathrm{vib}} - \omega_{\mathrm{i}} + \omega_{\mathrm{s}} - i\gamma)(\omega_{\mathrm{e_2g}} - \omega_{\mathrm{i}} - i\Gamma_{\mathrm{eg}})} \tag{67}$$

$$\chi^{(3)}_{\mathrm{G8}} = \frac{N}{\varepsilon_0 \hbar^3} \sum_{\mathrm{e_1,e_2}} \frac{(-1)^3 \mu_{\mathrm{ve_2}} \mu_{\mathrm{e_2g}} \mu_{\mathrm{ge_1}} \mu_{\mathrm{e_1v}}}{(-\omega_{\mathrm{e_1v}} - \omega_{\mathrm{i}} - i\Gamma_{\mathrm{ev}})(\omega_{\mathrm{vib}} - \omega_{\mathrm{i}} + \omega_{\mathrm{s}} - i\gamma)(-\omega_{\mathrm{e_2v}} - \omega_{\mathrm{i}} - i\Gamma_{\mathrm{ev}})} \tag{68}$$

$$\chi^{(3)}_{\mathrm{L5}} = \frac{N}{\varepsilon_0 \hbar^3} \sum_{\mathrm{e_1,e_2}} \frac{\mu_{\mathrm{ve_1}} \mu_{\mathrm{e_1g}} \mu_{\mathrm{ge_2}} \mu_{\mathrm{e_2v}}}{(\omega_{\mathrm{e_1v}} - \omega_{\mathrm{s}} - i\Gamma_{\mathrm{ev}})(-\omega_{\mathrm{vib}} + \omega_{\mathrm{i}} - \omega_{\mathrm{s}} - i\gamma)(\omega_{\mathrm{e_2v}} - \omega_{\mathrm{s}} - i\Gamma_{\mathrm{ev}})} \tag{69}$$

$$\chi^{(3)}_{\mathrm{L6}} = \frac{N}{\varepsilon_0 \hbar^3} \sum_{\mathrm{e_1,e_2}} \frac{(-1) \mu_{\mathrm{ve_1}} \mu_{\mathrm{e_1g}} \mu_{\mathrm{ge_2}} \mu_{\mathrm{e_2v}}}{(\omega_{\mathrm{e_1v}} - \omega_{\mathrm{s}} - i\Gamma_{\mathrm{ev}})(-\omega_{\mathrm{vib}} + \omega_{\mathrm{i}} - \omega_{\mathrm{s}} - i\gamma)(-\omega_{\mathrm{e_2g}} - \omega_{\mathrm{s}} - i\Gamma_{\mathrm{eg}})} \tag{70}$$

$$\chi^{(3)}_{\mathrm{L7}} = \frac{N}{\varepsilon_0 \hbar^3} \sum_{\mathrm{e_1,e_2}} \frac{\mu_{\mathrm{ve_1}} \mu_{\mathrm{e_1g}} \mu_{\mathrm{ge_2}} \mu_{\mathrm{e_2v}}}{(\omega_{\mathrm{e_1v}} + \omega_{\mathrm{i}} - i\Gamma_{\mathrm{ev}})(-\omega_{\mathrm{vib}} + \omega_{\mathrm{i}} - \omega_{\mathrm{s}} - i\gamma)(\omega_{\mathrm{e_2v}} - \omega_{\mathrm{s}} - i\Gamma_{\mathrm{ev}})} \tag{71}$$

$$\chi^{(3)}_{\mathrm{L8}} = \frac{N}{\varepsilon_0 \hbar^3} \sum_{\mathrm{e_1,e_2}} \frac{(-1) \mu_{\mathrm{ve_1}} \mu_{\mathrm{e_1g}} \mu_{\mathrm{ge_2}} \mu_{\mathrm{e_2v}}}{(\omega_{\mathrm{e_1v}} + \omega_{\mathrm{i}} - i\Gamma_{\mathrm{ev}})(-\omega_{\mathrm{vib}} + \omega_{\mathrm{i}} - \omega_{\mathrm{s}} - i\gamma)(-\omega_{\mathrm{e_2g}} - \omega_{\mathrm{s}} - i\Gamma_{\mathrm{eg}})} \tag{72}$$

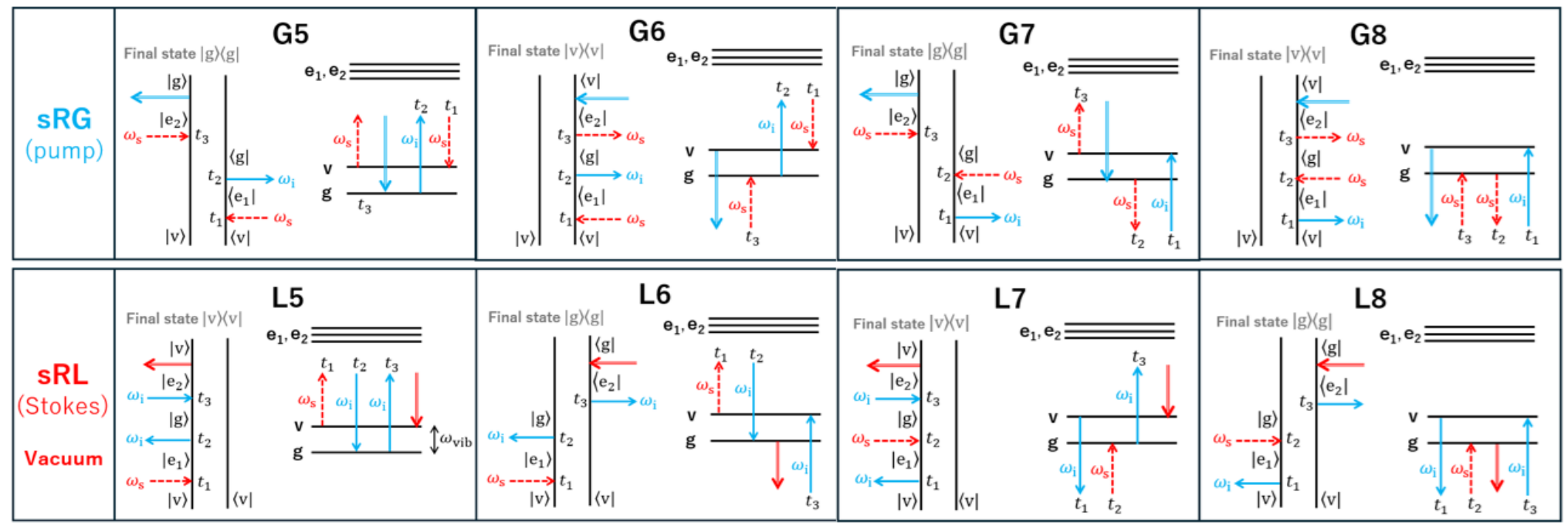

Fig. 4. Double-sided Feynman diagrams and energy level diagrams describing G5–G8 in sRG and L5–L8 in sRL.

The aforementioned eight unobservable terms cannot be detected during the measurement of Stokes spontaneous Raman scattering. In this specific process, only the incident field with an angular frequency of $\omega_{\mathrm{i}}$ irradiates the material, and prior to the interaction, no real photons exist at $\omega_\mathrm{s}$; only the vacuum field is present at this frequency. It is important to note that in coherent Raman scattering—where both the $\omega_\mathrm{i}$ and $\omega_\mathrm{s}$ fields externally irradiate the material—all terms (L1–L8 and G1–G8) contribute to the process if the initial population exists. Conversely, in spontaneous Raman scattering, only the observable terms (L1–L4 and G1–G4) are physically realized, and the unobservable terms (L5–L8 and G5–G8) are strictly prohibited due to the following relations.

$$\text{sRG (unobservable terms):}\quad |E_\mathrm{i}|^2 - 2\mathrm{Im}\left\{\chi^{(3)}_\mathrm{SRG}\right\}|E_\mathrm{i}|^2|V_\mathrm{s}|^2 \tag{73}$$

$$\text{sRL (unobservable terms):}\quad |V_\mathrm{s}|^2 - 2\mathrm{Im}\left\{\chi^{(3)}_\mathrm{SRL}\right\}|E_\mathrm{i}|^2|V_\mathrm{s}|^2 \tag{74}$$

In these expressions, the roles of the pump field and the vacuum field are reversed compared to Eqs. (16) and (17). The sRL in unobservable terms represents the loss in the vacuum field, but the vacuum field is the ground state of the electromagnetic field and does not fall to a lower energy state, which leads to the strict prohibition rule. The sRG in unobservable terms represents the gain of the incident field, but this phenomenon is also prohibited because there is no energy supply. Consequently, even if an initial population exists in the state $|\mathrm{v}\rangle\langle\mathrm{v}|$, the processes corresponding to terms L5–L8 and G5–G8 cannot occur.

## C. Negative-frequency-assisted quantum mechanical approach

Next, we generalize the KHD formula. As illustrated in Fig. 1 and Eq. (1), the traditional KHD theory consists of two primary components: the rotating term (①) and the counter-rotating term (②). Furthermore, as depicted in Fig. 5, we propose a quantum mechanical formulation by introducing two additional terms, ③ and ④, termed the "negative-frequency rotating" and "negative-frequency counter-rotating" terms, respectively. We designate them as "negative-frequency" because they formally correspond to transitions from unpopulated vibrational states, initiated by an interaction with a negative-frequency Stokes field that leaves population at the vibrational state and takes away population from the ground state: $-\omega_\mathrm{s} - (-\omega_\mathrm{i}) = \omega_\mathrm{vib}$.

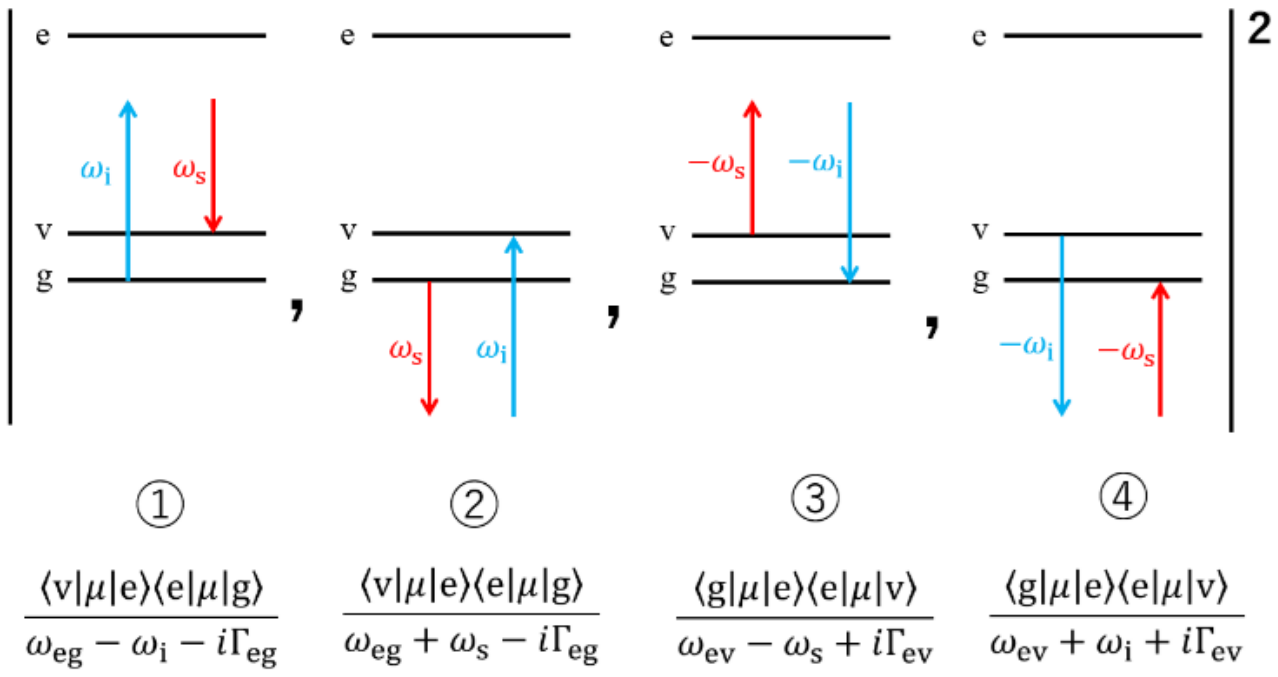

Fig. 5. Four terms involved in quantum mechanical formulation: ① and ② are the rotating and counter-rotating terms, respectively; and ③ and ④ are the negative-frequency-rotating and negative-frequency-counter-rotating terms, respectively. In ③ and ④ that treat the negative frequency, the signs of damping factors become positive.

To bridge this negative-frequency-assisted quantum mechanical approach and the nonlinear optical approach, we replace $|① + ②|^2$ with the following formula:

$$\left|①, ②, ③, ④\right|^2 \equiv M = \begin{pmatrix} ①^* & ②^* \\ -③^* & -④^* \end{pmatrix} \otimes \begin{pmatrix} -③ & -② \\ ① & ④ \end{pmatrix}, \tag{75}$$

where a symbol ⊗ represents Kronecker product. The calculation yields a four-by-four matrix with sixteen distinct components.

$$M = \begin{pmatrix} -①^*③ & -①^*② & -②^*③ & -②^*② \\ ①^*① & ①^*④ & ②^*① & ②^*④ \\ ③^*③ & ③^*② & ④^*③ & ④^*② \\ -③^*① & -③^*④ & -④^*① & -④^*④ \end{pmatrix} \quad (76)$$

As in the previous section (DISCUSSION A), we take the real part of each term and multiply it by the coefficient A defined in Eq. (53a) as shown in Table 1. Crucially, the real part of every single term in this matrix, multiplied by A, is found to be in exact analytical agreement with the derivations from the nonlinear susceptibility framework, sharing identical coefficients. Here we used complex conjugate relationships $\mathrm{Re}\{① ③^*\} = \mathrm{Re}\{①^* ③\}$ etc. and the frequency relations $\omega_{\mathrm{ev}} - \omega_{\mathrm{s}} = \omega_{\mathrm{eg}} - \omega_{\mathrm{i}}$ and $\omega_{\mathrm{ev}} + \omega_{\mathrm{i}} = \omega_{\mathrm{eg}} + \omega_{\mathrm{s}}$, and the assumption of the damping factor $\Gamma_{\mathrm{ev}} = \Gamma_{\mathrm{eg}} = \Gamma$ is applied. The mapping rules in Table 1 established by Eq. (75), each term corresponds uniquely to one of the sixteen susceptibility terms (L1–L4 and G1–G4 in Fig. 2, and G5–G8 and L5–L8 in Fig. 4). The minus and plus signs express the Raman loss and Raman gain, respectively.

Table 1. Correspondence table between the negative-frequency-assisted quantum mechanical formula and the nonlinear susceptibilities for L1–8 and G1–8.

| | | | | |
|---|---|---|---|---|
| **Observable terms** | **L1**<br>$-A\mathrm{Re}\{①^*③\}$ | **L2**<br>$-A\mathrm{Re}\{①^*②\}$ | **L3**<br>$-A\mathrm{Re}\{②^*③\}$ | **L4**<br>$-A\mathrm{Re}\{②^*②\}$ |
| | **G1**<br>$A\mathrm{Re}\{①^*①\}$ | **G2**<br>$A\mathrm{Re}\{①^*④\}$ | **G3**<br>$A\mathrm{Re}\{②^*①\}$ | **G4**<br>$A\mathrm{Re}\{②^*④\}$ |
| **Unobservable terms** | **G5**<br>$A\mathrm{Re}\{③^*③\}$ | **G6**<br>$A\mathrm{Re}\{③^*②\}$ | **G7**<br>$A\mathrm{Re}\{④^*③\}$ | **G8**<br>$A\mathrm{Re}\{④^*②\}$ |
| | **L5**<br>$-A\mathrm{Re}\{③^*①\}$ | **L6**<br>$-A\mathrm{Re}\{③^*④\}$ | **L7**<br>$-A\mathrm{Re}\{④^*①\}$ | **L8**<br>$-A\mathrm{Re}\{④^*④\}$ |

As previously discussed, however, the unobservable processes (L5–L8 and G5–G8) remain physically forbidden in this scattering regime; consequently, only the observable terms (L1–L4 and G1–G4) actively contribute to the process. It is notable that while the terms consisting only of the negative frequency terms ③ and ④ belong to the unobservable phenomena, the half of the cross terms with ① and ② contributes to the observable processes.

## D. Hilbert space versus Liouville space

Fundamentally, the KHD theory is formulated in Hilbert space, whereas the calculation of the nonlinear optical susceptibility relies on the density matrix defined in Liouville space [12−14]. Although the traditional KHD formula lacks contributions from the L1, G2, L3, and G4 terms, the negative-frequency-assisted quantum mechanical framework presented here completely accounts for all concurrent observable terms (L1–L4 and G1–G4). Conversely, the macroscopic susceptibility formula in nonlinear optics intrinsically encompasses all of these terms. Moreover, it provides the flexibility to set an arbitrary damping factor ($\gamma > +0$), which dictates the Raman spectral bandwidth. Consequently, the nonlinear optical formalism mathematically subsumes the negative-frequency-assisted quantum mechanical approach.

## IV. CONCLUSION

In conclusion, we have successfully resolved the long-standing theoretical divide between quantum mechanics and macroscopic nonlinear optics in the description of Raman scattering. By introducing the concept of the quantum vacuum field into the framework of nonlinear optics, we revealed that the fundamental bridge between these two realms lies in the interference at the detector between the signal emitted from molecules and the vacuum field itself that stimulates the molecules. Furthermore, we generalized the traditional KHD theory by incorporating the negative frequency contribution and the resulting essential unobservable terms. Crucially, we proved that the nonlinear optical susceptibility—formulated in Liouville space with an arbitrary damping factor—inherently encompasses this negative-frequency-assisted quantum mechanical approach defined in Hilbert space, leading to a perfect analytical agreement in the observed spontaneous Raman intensities. Beyond completing this theoretical foundation, our unified approach predicts a groundbreaking phenomenon: "spontaneous Raman loss". This discovery not only redefines our understanding of spontaneous optical processes but also opens new avenues for exploring vacuum-seeded light-matter interactions (e.g. spontaneous Rayleigh loss) and developing novel spectroscopic techniques.

## SUPPLEMENTARY MATERIAL

The supplementary material encompasses a description of classical theory to support the conclusion in the main text.

## ACKNOWLEDGMENTS

This study was financially supported by Japan Society for the Promotion of Science KAKENHI Grant Number 24K03251 (Grant-in-Aid for Scientific Research [B]).

## DATA AVAILABILITY

The data that supports the findings of this study are available within the article and its supplementary material.

# SUPPLENTARY MATERIAL

# Bridging quantum mechanics and nonlinear optics in Raman scattering

Naoki Fukutake[1,2*] and Hideaki Kano[3]

[1]*Nikon Corporation, 471 Nagaodai-cho, Sakae-ku, Yokohama-city, Kanagawa 244-8533, Japan*
[2]*Institute of Pure and Applied Science, University of Tsukuba, Tennodai 1-1-1, Tsukuba, Ibaraki 305-8573 Japan.*
[3]*Department of Biosciences and Informatics, Faculty of Science and Technology, Keio University,3-14-1 Hiyoshi, Kohoku-ku, Yokohama, Kanagawa 223-8522, Japan.*

*Naoki.Fukutake@nikon.com

## CLASSICAL TEORY

For comparison, we calculate the intensity of spontaneous Raman scattering using classical theory. We consider the following situation: when an excitation plane wave is incident onto target molecules, a scattered spherical wave is adiated from the target. The energy of the spontaneous Raman-scattered light that flows through a unit solid angle in a unit time, $I_{\mathrm{s}}$, can be expressed as [1]

$$I_{\mathrm{s}} = \frac{\omega_{\mathrm{s}}^4}{c^4}\alpha^2 I_{\mathrm{i}}\,, \tag{1}$$

where $\alpha$ represents the Raman scattering component of polarizability, $c$ denotes the speed of light, $\omega_{\mathrm{s}}$ stands for an angular frequency of scattered light, and $I_{\mathrm{i}}$ is the energy of the incident plane wave that flows through a unit area in a unit time. If the relation between the photon flux $F_{\mathrm{i}}$ and energy flux $I_{\mathrm{i}}$ for the incident light, $I_{\mathrm{i}} = \hbar\omega_{\mathrm{i}}F_{\mathrm{i}}$, where $\hbar$ denotes the reduced Planck constant and $\omega_{\mathrm{i}}$ stands for an angular frequency of the incident light, and the relation between the photon flux $F_{\mathrm{s}}$ and energy flux $I_{\mathrm{s}}$ for the Raman scattered light, $I_{\mathrm{s}} = \hbar\omega_{\mathrm{s}}F_{\mathrm{s}}$, are considered, Eq. (1) becomes

$$F_{\mathrm{s}} = \frac{\omega_{\mathrm{i}}\omega_{\mathrm{s}}^3}{c^4}\alpha^2 F_{\mathrm{i}}\,. \tag{2}$$

Thus, by introducing the quantum picture of photon flux into the classical theory, the first-and-third-power law $\omega_{\mathrm{i}}\omega_{\mathrm{s}}^3$ can be derived [2].